\documentclass[reprint, superscriptaddress, amsmath, amssymb, aip, jap,]{revtex4-1} 
\usepackage[allow-number-unit-breaks]{siunitx} 
\usepackage{graphicx}
\usepackage{amsmath}
\usepackage{dcolumn}
\usepackage{bm}
\usepackage[mathlines]{lineno}
\usepackage[breaklinks, bookmarks=false]{hyperref}
\hypersetup{
    pdfstartview=FitH,
    pdfnewwindow=true,      
    colorlinks=true,       
    linkcolor=blue,          
    citecolor=blue,        
    filecolor=blue,      
    urlcolor=blue           
}

\newcommand{\Ohm}{\ensuremath{\ohm}}						
					
\newcommand{\Hz}{\ensuremath{\textup{Hz}}}						
					
\newcommand{\PSD}{\ensuremath{\textup{V}_{\textup{rms}}^2/\textup{Hz}}} 		
\newcommand{\VSD}{\ensuremath{\textup{nV}/\sqrt{\textup{Hz}}}}

\newcommand{\Mean}[1]{\overline{#1}}		 		
\newcommand{\MeanSquare}[1]{\overline{{#1}^2}} 		
				
\newcommand{\StDev}[1]{\sigma_{#1}}

\newcommand{\Spec}{S}						
\newcommand{\SN}{S_N}						
\newcommand{\fr}{f_r}						
\newcommand{\fs}{f_s}						
\newcommand{\td}{t_d}						
\newcommand{\tm}{t}							
\newcommand{\freq}{f}						
\newcommand{\w}{\omega}						
\newcommand{\p}{\varphi}					
\newcommand{\e}{\varepsilon}				
\newcommand{\de}{\delta\varepsilon}			
\newcommand{\x}{x}							

\begin{document}
\title{Cross-spectrum Analyzer for Low Frequency Noise Analysis }
\author{Xing Zhong}
 \affiliation{MINT Center, University of Alabama, Tuscaloosa AL 35487}
 \affiliation{Department of Physics and Astronomy, University of Alabama, Tuscaloosa AL 35487}
\author{Sahar Keshavarz}
 \affiliation{MINT Center, University of Alabama, Tuscaloosa AL 35487}
 \affiliation{Department of Physics and Astronomy, University of Alabama, Tuscaloosa AL 35487}
 \author{Josh Jones}
 \affiliation{MINT Center, University of Alabama, Tuscaloosa AL 35487}
 \affiliation{Department of Physics and Astronomy, University of Alabama, Tuscaloosa AL 35487}
\author{Claudia Mewes}
 \affiliation{MINT Center, University of Alabama, Tuscaloosa AL 35487}
 \affiliation{Department of Physics and Astronomy, University of Alabama, Tuscaloosa AL 35487}
\author{Patrick R. LeClair}
 \affiliation{MINT Center, University of Alabama, Tuscaloosa AL 35487}
 \affiliation{Department of Physics and Astronomy, University of Alabama, Tuscaloosa AL 35487}

\begin{abstract}
 The design and performance of a sensitive and reliable cross-correlation spectrum analyzer for studying low frequency transport noise is described in detail. The design makes use of common PC-based data acquisition hardware and preamplifiers to acquire time-based data, along with software we have developed to compute the cross-correlation and noise spectral density.  
 The impedance of device under test may cover four decades from ${100\,{\Omega}}$ to ${1\,{\mathrm{M}\Omega}}$. By utilizing a custom developed signal processing program, this system is tested to be accurate and efficient for measuring voltage noise as low as $\sim\!{10^{-19}\,\PSD}$ from ${0.001\,}$Hz to ${100\,}$kHz within one day's averaging time, comparable with more expensive hardware solutions (bandwidth in real measurements may be limited by the sample impedance and stray capacitance).
 The time dependence of measurement sensitivity is discussed theoretically and characterized experimentally to optimize between measuring time and accuracy.
 A routine for noise component analysis is introduced, and is applied for characterizing the noise spectra of metal and carbon film resistors, revealing an almost strict $1/$frequency dependence that may reflect an ensemble of random resistivity fluctuation processes with uniformly distributed activation energies. These results verify the general applicability of this analyzer for low level noise researches.
\end{abstract}
\maketitle

\section{Introduction}
 Noise spectral analysis is proven to be a useful method to study the carrier transport characteristics in various materials. In most noise spectra taken with a sample under bias, the low-frequency behavior is dominated by a power spectral density with a $1/$frequency dependence, known as ``$1/f$" or ``flicker" noise. The $1/f$ noise contribution, readily revealed by a constant current through the device, provides important information on device resistivity fluctuations\cite{Hooge1981} and is extremely useful when studying the transport mechanisms in semiconductors materials,\cite{Bess1956,Mihaila2000} magnetic tunneling junctions,\cite{Mazumdar2007,Guerrero2006}, p-n junctions,\cite{Kleinpenning1985,Akiba1997}, percolation networks,\cite{Rammal1985,Soliveres2007} and other areas.\cite{Podzorov2001,Kolek2007,Mantese1981}
 
In order to minimize the background noise introduced by the analyzer components, such as the preamplifiers, cross-correlation methods are routinely applied for noise measurements.\cite{Sampietro1999,Jonker1999}
 In this method, the noise voltage signal is \emph{simultaneously} acquired in two channels from two independent pairs of electrical contacts to the device. By calculating the cross-correlation spectrum of the two time-domain signals, the uncorrelated components will be eliminated and remaining contributions will be only the common-mode signal from the device itself. For conciseness, it is abbreviated as cross-spectrum throughout this article, and it is usually calculated by the following routine.\cite{Sampietro1999,Dicarlo2006}
 
This article is organized as follows: the mathematical background of the cross-spectrum approach for noise analysis is introduced in section II. Time-sensitivity of the measurements are explained in section III. The schematic of the system hardware and software are described along with system specifications and measurement limitations in section IV and V. Noise-component analysis is explained in section VI, and finally in Sections VII and VIII we present the measurement method and analysis of two types of noise: thermal and $1/f$ noise.. 

\section{Mathematical Background}

 Consider two continuous time domain signals $x(t)$ and $y(t)$. We can define their cross correlation function as:
 \[
 R_{xy}(t) = \bigl(x \star y\bigr)(t) \equiv \lim\limits_{T \rightarrow\infty} \int\limits_{-T/2}^{T/2} x(\tau)y^\ast (\tau-t) d\tau 
 \]
 where $^\ast$ denotes complex conjugation. According to the Wiener-Khinchin theorem, the power spectral density (PSD), $S_{xy}(\omega)$, is the Fourier transform ($\mathcal{F}$) of the cross correlation function:\cite{Hooge1981}
 \[
 S_{xy}(\omega) = \mathcal{F}\left[R_{xy}(t)\right] 
 \]
 \indent The cross-correlation theorem states that $\mathcal{F}\left[f\star g\right]\!=\!\bigl(\mathcal{F}\left(f\right)\bigr)^\ast \mathcal{F}\left(g\right)$, which gives
 \[
 S_{xy}(\omega) = \mathcal{F}\left[\left(x \star y\right)(t)\right] = \bigl(\mathcal{F}\left[x(t)\right]\bigr)^\ast
 \mathcal{F}\left[y(t)\right]
 \]
 If $x(t)$ and $y(t)$ have Fourier transforms $\tilde{x}(\omega)$ and $\tilde{y}(\omega)$,
 \[
 S_{xy}(\omega) =  \mathcal{F}\left[x(t)\right]^\ast \mathcal{F}\left[y(t)\right] = \tilde{x}^\ast(\omega)\tilde{y}(\omega)
 \]
 In the case of single-sided spectra (i.e., positive frequencies only), one must double the Fourier transform of both $x(t)$ and $y(t)$, giving:
 \begin{equation}
  S_{xy}(\omega) =  4\tilde{x}^\ast(\omega)\tilde{y}(\omega)
  \label{eqn:xcorrFormula}
 \end{equation}

In short, this means that two simultaneous voltage versus time measurements $x(t)$ and $y(t)$, combined with subsequent signal processing to determine $\tilde{x}^\ast(\omega)$ and $\tilde{y}(\omega)$, are sufficient to determine the noise power spectral density as a function of frequency. The above definitions are strictly valid only for square integrable signals; in the more general case of, e.g., wide-sense stationary processes, the Fourier transforms of $x(t)$ and $y(t)$ do not necessarily exist. In this case the cross correlation function can be formulated in terms of the expected value function rather than an indefinite integral, but the same essential conclusions hold. Similarly, the primary result is readily generalized to the case of discrete rather than continuous signals. In this article, we describe a combined hardware and software solution for computing the noise spectral density as a function of frequency from simultaneously acquired voltage versus time measurements. 

\section{Time-dependent Sensitivity} \label{sec:time-dep}

In order to properly determine the noise spectral density of a device under test (DUT), we should first consider the measurement problem in more detail. In general, a sampled time domain signal can be expanded as a discrete Fourier series:

\begin{equation}
v(t) = \sum\limits_{k = - \frac{N}{2}}^{\frac{N}{2}} {V({\w_k})} \label{eq:dfs}
\end{equation}

\noindent where ${\w_k} = 2k\pi {\fs}/N$, $\fs$ is the sampling frequency, and $N$ is the number of samples. Essentially, the voltage signals can be seen as the superposition of a series of sinusoidal waves of frequencies $\w_k$. The frequency component for each channel, $V(\w_k)$, can be decomposed in to separate contributions from intrinsic noise of the DUT, and the external noise including the environmental sources, such as $60\,$Hz multipliers from power lines and instrumental noise introduced by the hardware of the two channels during signal transmission and amplification: 

\begin{equation}
\begin{gathered}
  {V_{Ch1}}({\w _k}) = \dfrac{1}{2}(\de _k{e^{i({\w _k}t + {\p _k})}} + \de_{1,k}{e^{i({\w _k}t + {\p _{1,k}})}}) \hfill \\
  {V_{Ch2}}({\w _k}) = \dfrac{1}{2}(\de _k{e^{i({\w _k}t + {\p _k})}} + \de_{2,k}{e^{i({\w _k}t + {\p _{2,k}})}}) \hfill \\ 
\end{gathered} 
\end{equation}

\noindent where the $\de_k$ is the amplitude of the device noise and $\de_{1,k} ,\de_{2,k}$ represent the external noise in channel 1 and channel 2 at frequency $\omega_k$. Here $\p_k$, $\p_{1,k}$ and $\p_{2,k}$ represent the phases of the noise components, respectively. Considering positive frequencies only, $V(\w_k)$ is essentially equal to the half-sided Fourier transform of the noise signal divided by the number of samples. The cross-spectrum$\,$($\Spec$) is calculated from the real part of the product of $V_{Ch1}$ and the complex conjugate of $V_{Ch2}$\cite{Sampietro1999}, as shown in Eqn.~\ref{eqn:xcorrFormula}:

\begin{equation}
\begin{gathered}
\begin{aligned}
  S(\omega_k)  & = 4\Re ({V_{Ch1}}{V_{Ch2}}^*) \hfill\\
               & = \de_k ^2 + \de_k \de_{1,k} \cos(\p_k  - {\p _{1,k}}) + ... \hfill \\
               &... + \de_k \de _{2,k} \cos (\p_k - {\p _{2,k}}) + ...\hfill\\
               &... + \de_{1,k} \de_{2,k} \cos ({\p _{1,k}} - {\p _{2,k}}) \hfill \\ 
\end{aligned}
\end{gathered} 
\end{equation}

The spectral density $\Spec$ is then further divided by $2$ to obtain the root mean square power spectral density. One must note that $S(\omega_{k})$ is equal to the device's spectral density $\de_{k}^{2}$ only when the phase differences $\p_k-\p_{1,k}$, $\p_k-\p_{2,k}$, and $\p_{1,k}-\p_{2,k}$ are ${\pm}k\pi/2$, i.e., the contributions are mutually orthogonal. This can strictly only be achieved by measuring the cross-spectrum for an infinite duration of time. Practically speaking, only a spectrum of limited bandwidth over a limited measurement time can be determined, and therefore the expected spectrum is estimated by the averaging of $N$ independently acquired spectra$\,$($\Mean{\SN}$). Knowing the spectral frequency resolution$\,$($\fr$) is the reciprocal of sampling time, the total averaging time is then related to the number of averages and spectral resolution by $\tm = N/{\fr}$. The uncertainty in $\Mean{\SN}$ as a function of averaging time can then be readily determined:

\begin{equation}
\StDev{\Mean{\SN}}  = \frac{{{\StDev{\Spec}}}}{{\sqrt N }} = \frac{{{\StDev{\Spec}}}}{{\sqrt {\tm\cdot{\fr}} }}
\label{eqn:SdVsAveragingTime}
\end{equation}

\noindent where $\StDev{\Mean{\SN}}$ is the standard deviation of the averaged spectrum for a collection of $N$ measurements of $S$, each with standard deviation $\StDev{\Spec}$. When system background noise is much higher than the device noise $\,$($\de \ll \de_1 \approx \de_2$),
$\StDev{\Spec}$ can be experimentally measured by averaging the absolute value of $\Spec$ on a short circuited device:

\begin{equation}
\begin{gathered}
\begin{aligned}
  {\StDev{\Spec}} & = \Mean { | \de \de_1 \cos (\p  - {\p _1}) + \de \de_2 \cos (\p  - \p_2 ) + ...}  \hfill \\
               & \Mean {... + \de_1 \de_2 \cos (\p - \p _2) |} \hfill\\
               &\approx \Mean {|\de_1 \de_2 \cos (\p  - \p_2)|}  = \Mean {|{\Spec_{Short}}|}  \hfill \\
\end{aligned} 
\end{gathered} 
\end{equation}

\section{System Schematic}

\begin{figure}[h]
\begin{center}
  \includegraphics[width=0.95\columnwidth]{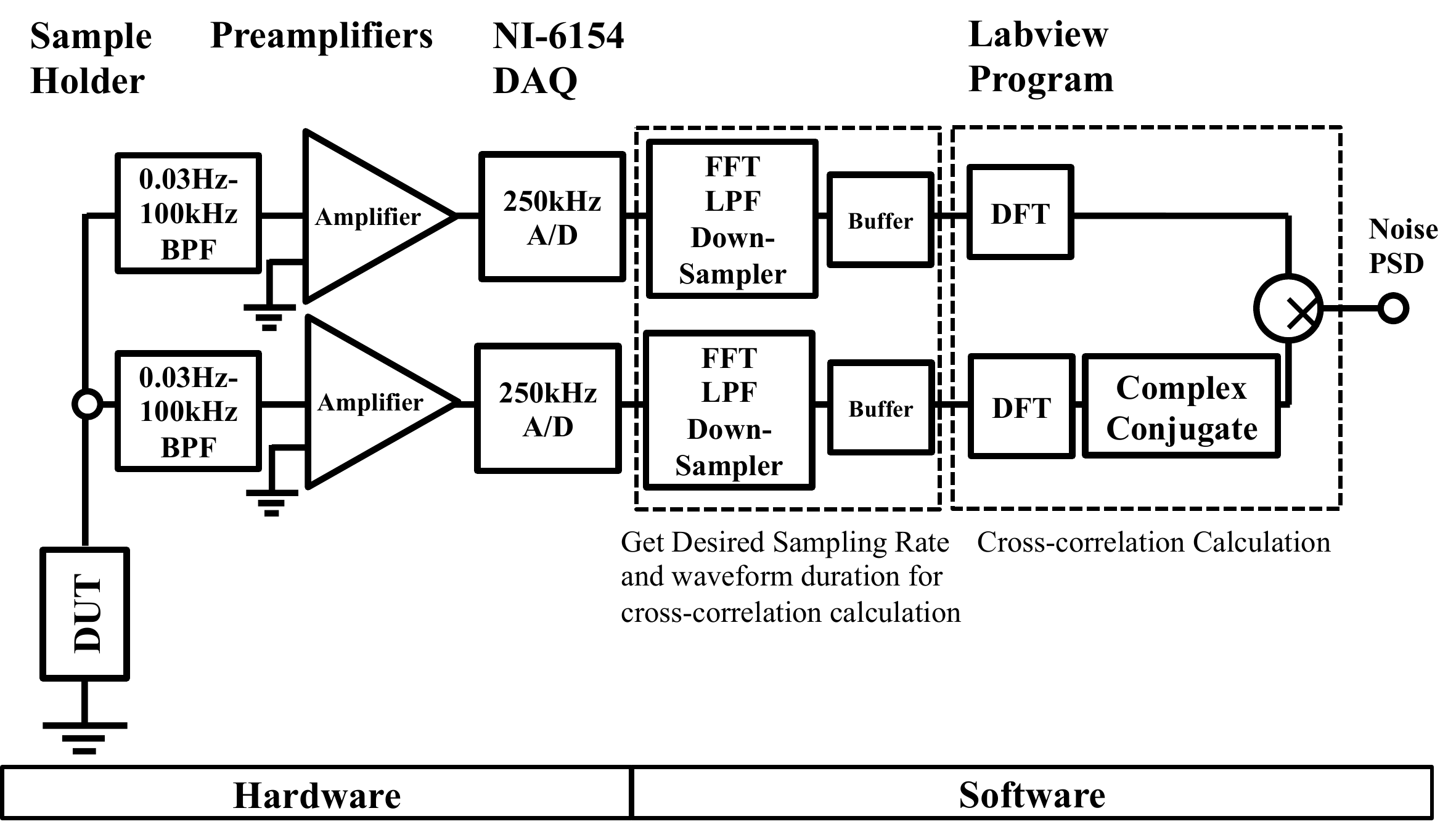}
\end{center}
  \caption{ Schematics of the building blocks of the cross-spectrum analyzer. (DUT = Device under test, BPF = Band-pass filter, AMP = amplifier, DAQ = Data Acquisition Card, AA = Anti-alias, FFT = Fast Fourier transform, Conj. = Complex Conjugate.)\label{fig:schematic} }
\end{figure}

The schematic diagram of our analyzer is illustrated in Fig.~\ref{fig:schematic}. The noise signal (voltage versus time) from the device under test is sampled through two independent pairs of contacts simultaneously. Each signal is then fed through the internal band-pass filter of two independent low-noise preamplifiers for dc blocking and anti-aliasing purposes. We have used both EG\&G 113 and Signal Recovery 5113\cite{5113} preamplifiers, and found the results to be essentially identical. The comparison was made for the primary reason that the EG\&G preamplifiers are no longer readily available, and we have found the Signal Recovery preamplifiers to be a straightforward ``drop in'' replacement. The critical factors in preamplifier selection we have found, other than having the lowest possible noise floor, are: (1)  ability to run on battery power, for at least 12 hours, (2) ability to turn off the digital display (to increase battery life and avoid picking up extraneous signals from the display), (3) internal band-pass filtering, and (4) adjustable gain. Of these factors, the first two appear to be the most crucial.

The subsequently amplified signals are sampled by a National Instruments 6154 16-bit channel-to-channel isolated data acquisition card (DAQ)\cite{NI}, running on continuous mode with sampling rate of $f_{s}\!=\!\SI{250}{\kilo\Hz}$. In this case, the channel-to-channel isolation and simultaneous sampling capabilities were the primary selection criteria, along with a sufficiently high sampling rate to allow measurements to $\sim\!100\,$kHz. Given the resolution of 16 bits and an input range of $1\,$V, this implies a minimum voltage increment due to analog-to-digital quantization of $\sim\!15\,\mu\text{V}$ without any averaging. Given a preamplifier gain of $10^3$ (for example; typically gains of $10^3$ or $10^4$ are used), the minimum measurable voltage with one channel is then of order $v_{\text{min}}\!\sim\!15\,\text{nV}$. For thermal (Johnson-Nyquist) noise, for a given measurement bandwidth $\Delta f$ the rms noise voltage is given by $v_n\!=\!\sqrt{4k_BTR\Delta f}$, where $k_B$ is Boltzmann's constant and $R$ is the resistance. A minimum measurable noise voltage of $v_{\text{min}}$ thus implies a minimum measurable resistance with a single-channel measurement. Over a $100\,$kHz bandwidth, this gives $R_{\text{min}}\!\sim\!0.1\,\Omega$. This is approximately four orders of magnitude below the observed single-channel noise floor in our system, indicating that discretization effects are not playing a significant role in our system.

\begin{figure}[h!]
\begin{center}
\includegraphics[width=0.95\columnwidth]{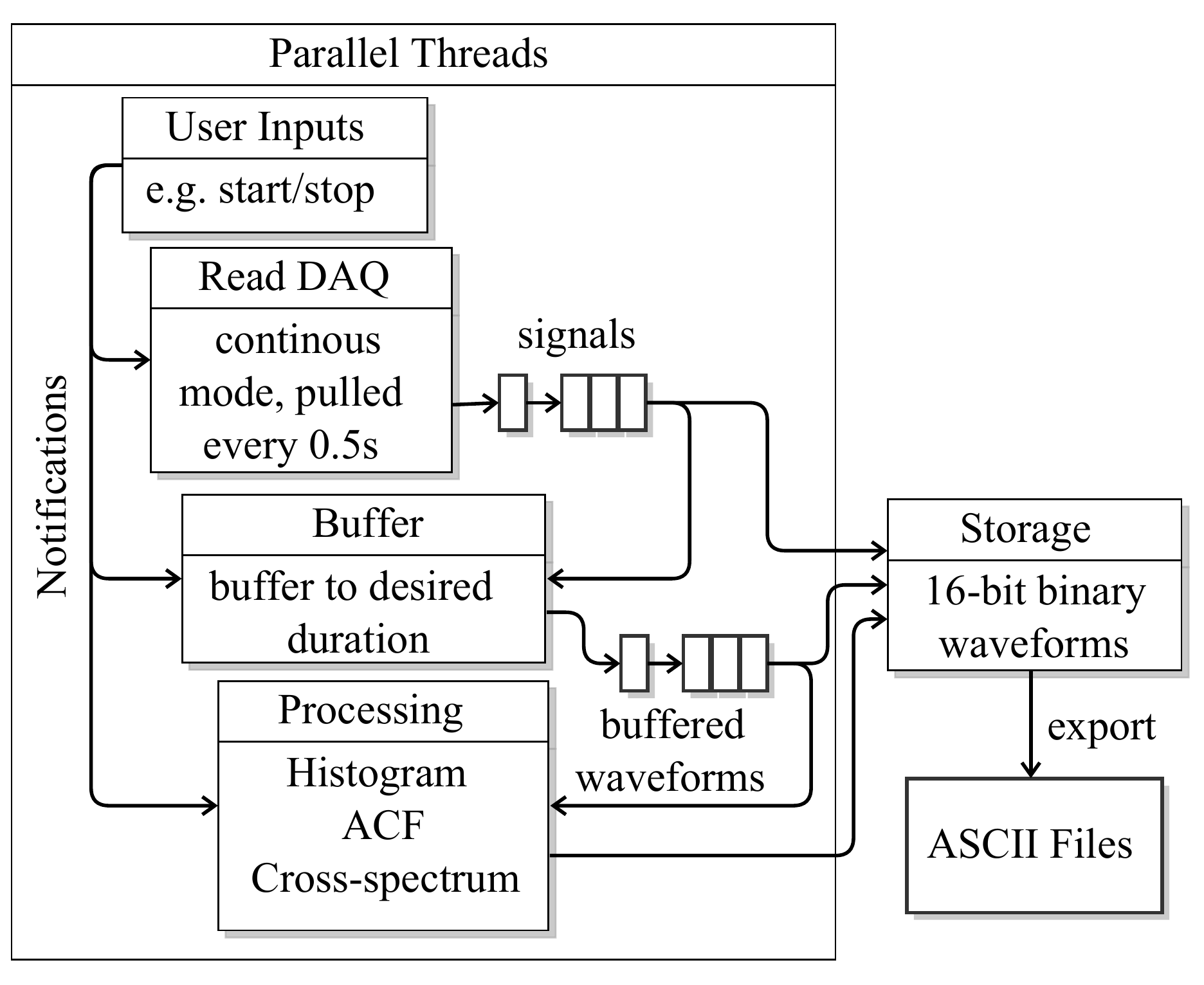}
\end{center}
\caption{Block diagram of the signal processing program with a producer/consumer design pattern.
\label{fig:prog_block}}
\end{figure}

A custom-developed Labview program buffers the acquired waveforms for a certain duration $\td$ to obtained the desired spectral solution ($\fr=1/\td$), applies an anti-alias filter if highest frequency of interest is lower than half the sampling frequency $f_{s}/2$, computes power spectral density, and then averages measured cross-spectra over time to improve sensitivity. The entire measurement process is automated, including the ability to run multiple measurements back-to-back for pre-defined sampling times. This program uses a producer/consumer design pattern\cite{P-C-Pattern} to realize real-time sampling, i.e.\ the spectrum is calculated at the same time when the waveforms are being sampled, so there is no delay time between each measurement for signal processing (Fig.~\ref{fig:prog_block}). Thanks to this software implementation, no extra hardware component is needed beyond the relatively standard amplifiers and DAQ, thus reducing the cost of the system. While the use of Labview does entail a significant cost, the program could be readily implemented on a number of other platforms (e.g., Matlab). The choice of Labview was motivated primarily by the fact it was already available, it is relatively easy to use, and many students are already familiar with it. This reflects the overall design philosophy: the system can be built using hardware and software which are either already available, or readily repurposed for other experiments if need be.

\section{Noise Floor}

\begin{figure}[h!]
\begin{center}
  \includegraphics[width=0.95\columnwidth]{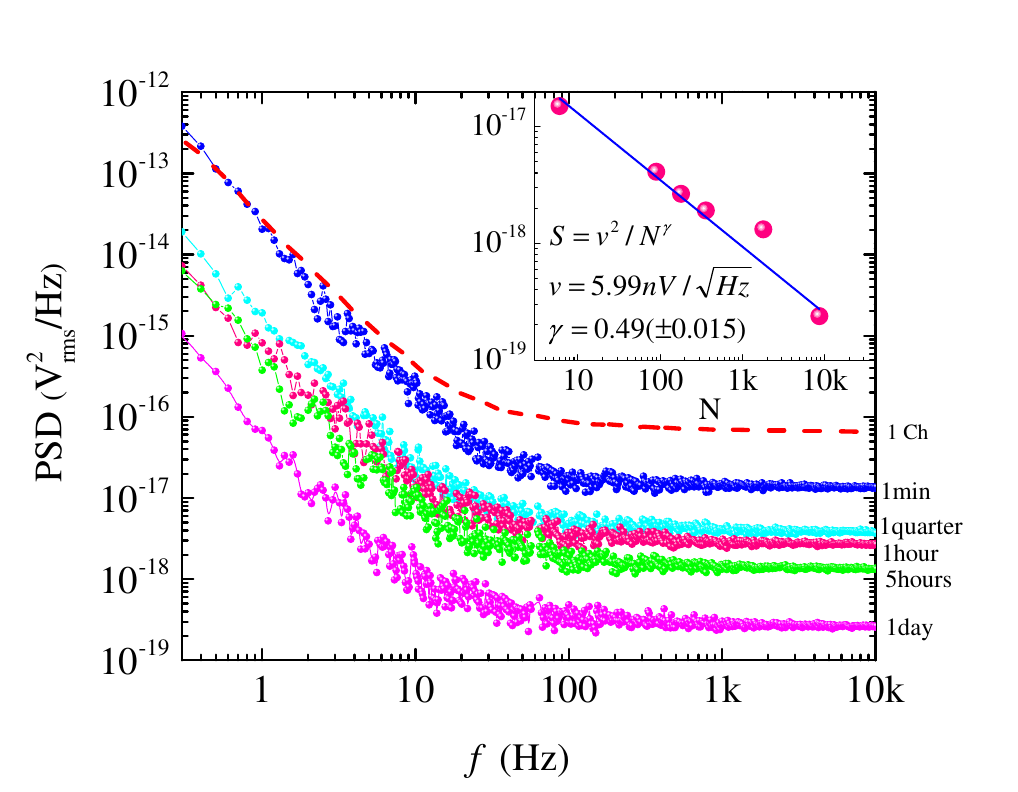}
\end{center}
  \caption{Measured {(dots)} cross-spectra PSD for a short circuit and calculated {(lines)} time dependence of sensitivity threshold for \SI{0.1}{\Hz} spectral resolution.  {Insert:} sensitivity threshold at \SI{1}{\kilo\Hz} as a function of number of averages, which fits to Eqn.~\ref{eqn:SdVsAveragingTime} remarkably well. 
  \label{fig:Time-sens} }
\end{figure}

The measured $\StDev{\Spec}$ for our system, whose general characteristic (smooth fit) is illustrated by the ``1ch'' curve in Fig.\ref{fig:Time-sens}, has a relatively complicated frequency dependence, because it combines noises from various sources in our system. Having $\StDev{\Spec}$, the time and averaging number dependence of the system sensitivity can be calculated from Eq.~\ref{eqn:SdVsAveragingTime}. Figure \ref{fig:Time-sens} shows the measured power spectral density (PSD), $\Mean{\SN}$, on a shorted device for various averaging times. Compared with system background noise, the sensitivity of the analyzer is improved by a hundred times by averaging the spectrum for one day ($\sim\!10$k averages), and by more than 10 times after averaging only $15\,$min. We note here that peaks in the noise spectra are still often observed at the line frequency of $60\,$Hz and its multiples due to external influences. While these peaks can be minimized and often eliminated by improved shielding and increased physical isolation of the experimental setup, for the purposes of analysis we have removed data in narrow windows ($\sim\!\pm\!5\,$Hz) around $60\,$Hz and its multiples. Typically these peaks are no larger than $10^{-14}\,\text{V}^2/\text{Hz}$ at $60\,$Hz, reduced by about two orders of magnitude at $120\,$ and $180\,$Hz, and may often be below the noise floor of the measurement. This brings to light the important task of shielding the measurement device and electronics from external sources of noise to the greatest degree possible, which we describe in Appendix~\ref{sec:shielding}.

\begin{figure}[h!]
\begin{center}
\includegraphics[width=0.95\columnwidth]{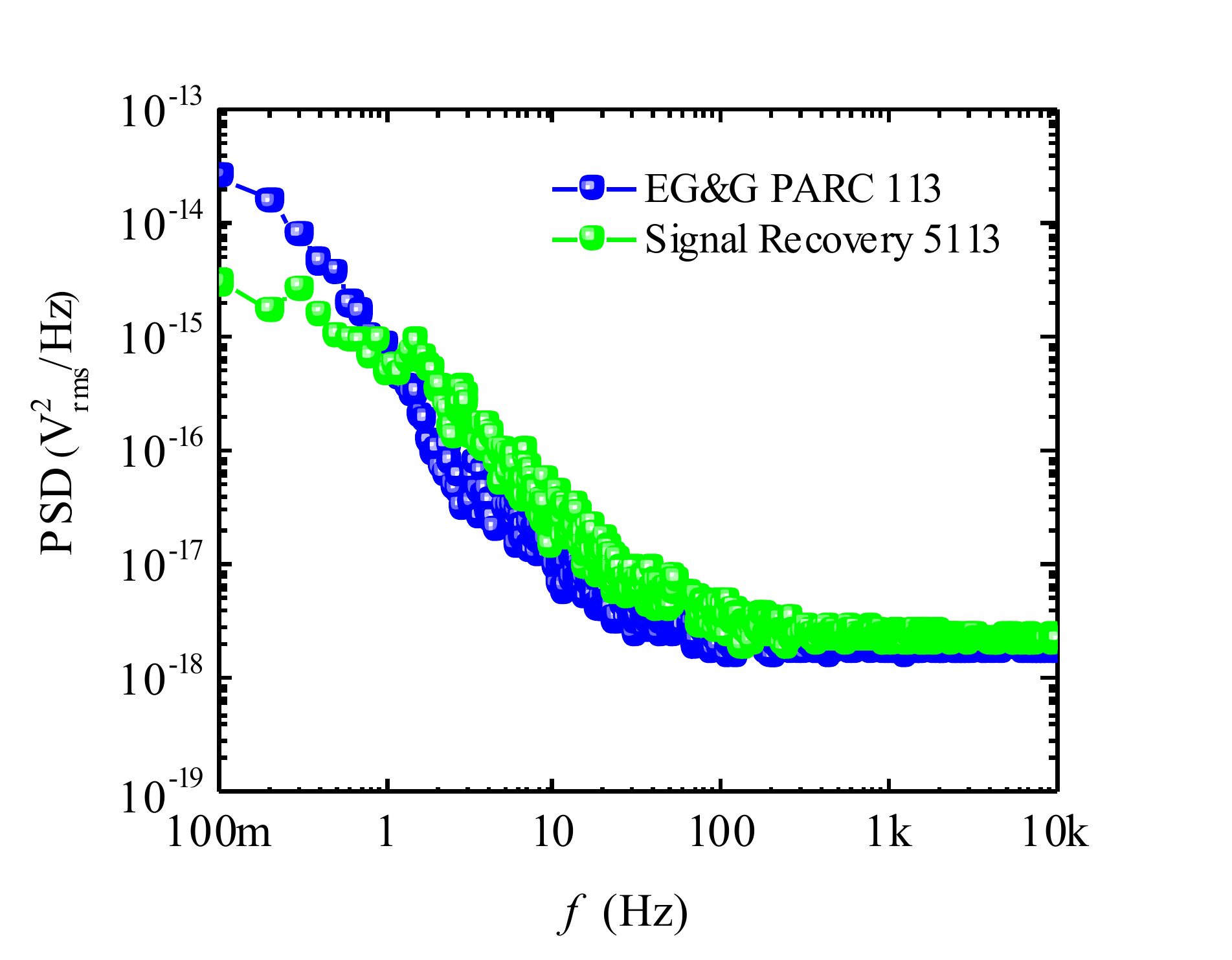}
\end{center}
  \caption{Cross-spectra for a short circuit measured for 1 hour with the EG\&G 113 and Signal Recovery 5113 preamplifiers, with and without a high-pass $RC$ filter set at $0.03\,$Hz.
  \label{fig:comparison} }
\end{figure}

The power spectral density at \SI{1}{\kilo\Hz} as a function of number of averages is plotted in the insert of Fig.~\ref{fig:Time-sens}, which fits the function $S=v^2/N^\gamma$ very well, with $v\!=\!5.99(\pm 0.05)\VSD$ and $\gamma\!=\!0.49(\pm 0.015)$. The measured $v$ is approximately the same as the rated background noise of our EG\&G 113 preamplifiers at \SI{1}{\kilo\Hz}, and $\gamma\!\sim\!0.5$ indicates the uncorrelated fluctuations follow normal distribution. In order to confirm that a more readily-available preamplifier may be used, we directly compared our EG\&G 113 preamplifiers with a Signal Recovery 5113 preamplifier. Figure~\ref{fig:comparison} shows the measured cross-spectra for a short circuit for 1 hour of measurement time for both amplifiers, with and without the use of the preamplifiers' high-pass $RC$ filters (set at $0.03\,$Hz for each amplifier, see Table~\ref{tab:RC}). The results from the two amplifiers are essentially indistinguishable. For the remainder of the paper, we report results obtained using the EG\&G 113 preamplifiers.

This result confirms that the residual correlation between the two channels is negligible, on the order of $10^{-19}\,\PSD$ after one day averaging time, verifying our assumption that the device noise and external noises in the two channels are mutually orthogonal. Another primary consideration is shielding the measurement device and electronics from external sources of noise to the greatest degree possible, we describe the physical layout of our system in Appendix \ref{sec:shielding}.

The maximum measurable frequency is set by the sampling rate through the Shannon-Nyquist theorem, while the minimum measurable frequency (frequency resolution of the measurements) can be set by buffering time adjustments. Combined with sampling and buffering considerations are the effects of the high- and low-pass filters in the preamplifiers themselves. These filters should be set to frequencies well outside the frequency range of interest if possible, their influence on the measured spectra is discussed in more detail in the following section. The high-pass filter is necessary in any event to eliminate any dc bias from reaching the preamplifiers.

\begin{figure}[h!]
\begin{center}
  \includegraphics[width=0.95\columnwidth]{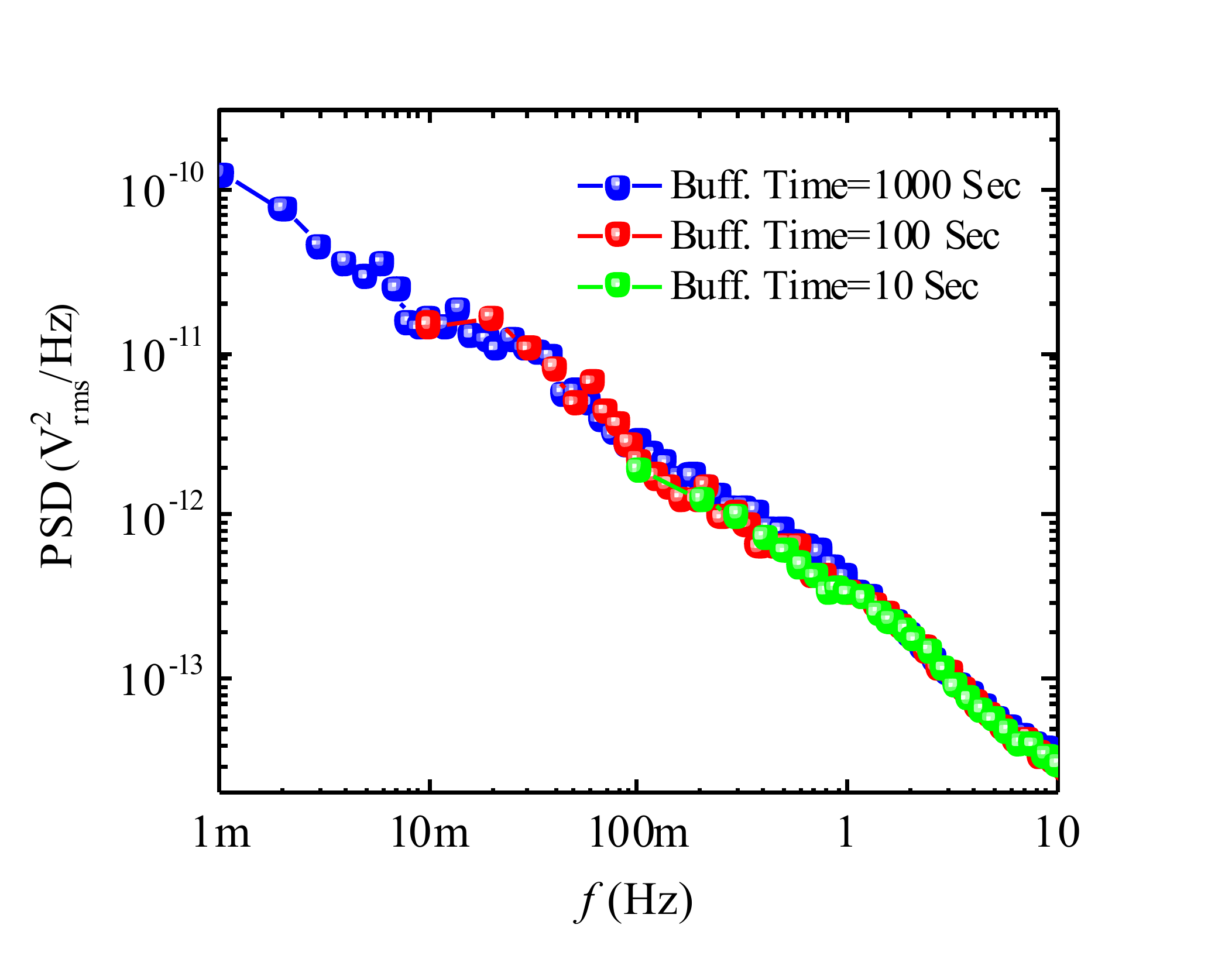}
\end{center}
  \caption{Cross-spectra for a $1\,\text{M}\Omega$ metal film resistor with a bias current of $1\,\mu\text{A}$ measured for different buffer times. 
  \label{fig:buffer} }
\end{figure}

As described in Sect.~\ref{sec:time-dep}, the frequency resolution of the spectrum (i.e., the lower frequency limit) can be set by adjusting the buffering times. Therefore, lower frequency measurements require longer buffering, and thus substantially longer averaging times in order to reduce the uncertainty in the spectrum for reliable measurements. Practically speaking, if it is determined that the experimental spectrum has reached the noise floor of the system at a particular minimum frequency, the buffering time can be adjusted to match that minimum frequency and the measurement time can be optimized accordingly. By software adjustment, the reliable frequency range can be set by adjusting the buffering time for lower frequency limit and adjusting the sampling rate for the upper frequency limit (and both for measurement time optimizations). For instance, frequency resolution of $1\,$mHz is possible by buffering time of $1000\,$s, while by setting the sampling rate to $2500\,$Hz, reliable spectra can be acquired up to $\sim\!1000\,$Hz, suitable for devices with impedance above $10\,\text{k}\Omega$ with $100\,$min total measurement time. 
Figure~\ref{fig:buffer} shows low frequency PSD of a $1\,\text{M}\Omega$ metal-film resistor, while applying $1\,\mu$A DC current as a probe of the $1/f$ noise (see Sect.~\ref{sec:1f}), with different buffer times and sampling rates adjusted for averaging time optimizations. It can be seen that PSD for all three buffer times and sampling rates are consistent. 

\section{Thermal Noise}

\begin{figure}[h!]
\begin{center}
  \includegraphics[width=0.95\columnwidth]{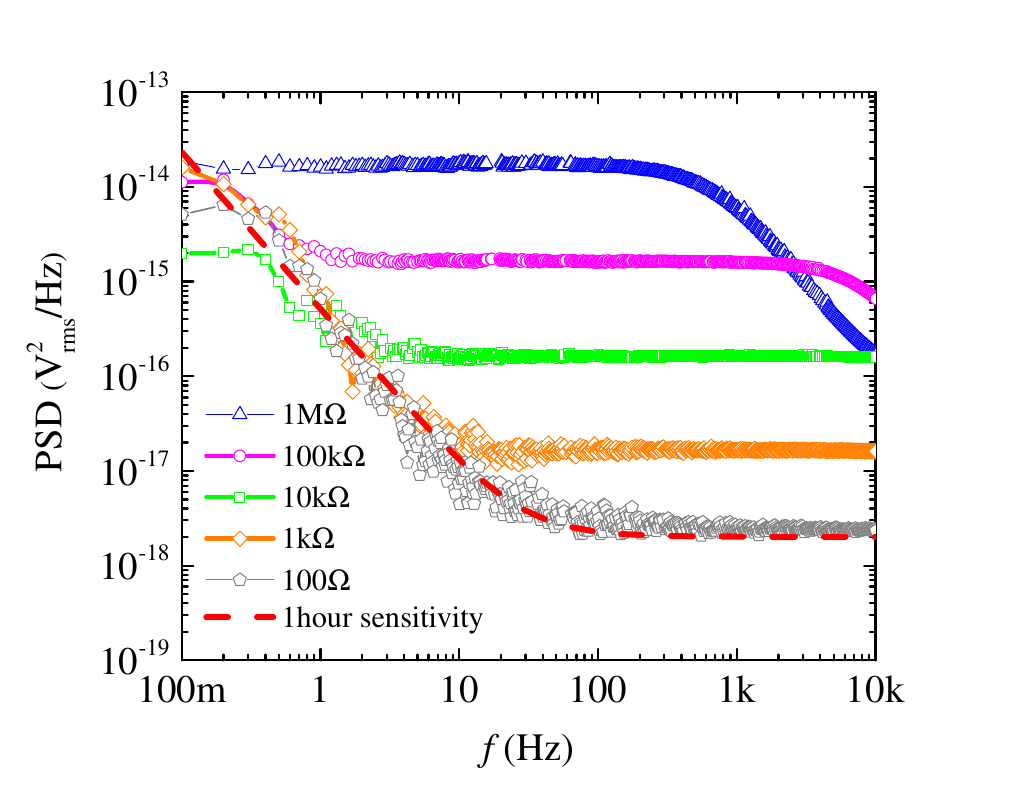}
\end{center}
  \caption{ One hour averages of thermal noise power spectra of various carbon composition resistors {(dots)} and the fits to theoretical values ($4k_{B}TR$) when considering stray capacitance effect {(lines)}. The sensitivity threshold for one hour averaging time and $0.1\Hz$ spectral resolution is marked by the dash line. \label{fig:White-noise} }
\end{figure}

The noise voltage measured on an ideal resistor is the Johnson-Nyquist noise, related to the thermal agitation of the carriers in a conductor. For an ideal resistive device of resistance $R$ at temperature $T$, for frequencies well below $\sim\!k_BT/h$ the power spectral density is given by $PSD\!=\!4k_{B}TR$ where $h$ is Planck's constant, $k_B$ is Boltzmann's constant, and $T$ the absolute temperature of the device. The PSD is independent of frequency, so-called ``white noise'', and proportional to the resistance of the device. To confirm our analysis for sensitivity, we measured the thermal noise power spectra at room temperature on carbon film resistors of various resistances $R$ for one hour averaging time(Fig.~\ref{fig:White-noise}) at room temperature $T\!\approx\!295\,$K. 

The measured white noise levels are very close to the expected values for thermal noise at room temperature, within the tolerance of the resistors, until the curves meet the measured sensitivity limit marked by the dashed line. The high frequency roll-off in the spectra is due two effects. First, the low-pass filter in the preamplifiers may roll off the spectrum at higher frequencies. Typically the filter is set at a sufficiently high frequency that this is outside the region of interest (typically, approximately 1 decade above the maximum frequency of interest). Second, and more importantly, for higher resistance devices an additional low pass filter is formed by the device resistance and the small stray capacitance $C$ between the testing port and ground (approximately $10\,$pF; Fig.~\ref{fig:eq-circuit}(a)). Either contribution has the effect of shaping the measured noise spectrum by a an RC filter response. Presuming the preamplifier low pass filter is set to a cutoff frequency $f_c$ outside the region of interest, we may consider only the device and stray capacitance and this results in multiplying the spectrum by $1/(1+(2{\pi}\freq R_{D} C)^2)$, where $R_D$ is the device resistance, and $C$ is the stray capacitance:

\begin{equation}
\MeanSquare {\de '}  = \frac{\MeanSquare{\de}}{{1 + {{(2\pi \freq {R_D} C)}^2}}}
\label{eqn:SpectrumShapedByLPF}
\end{equation}

\begin{figure}[h!]
\begin{center}
  \includegraphics[width=0.95\columnwidth]{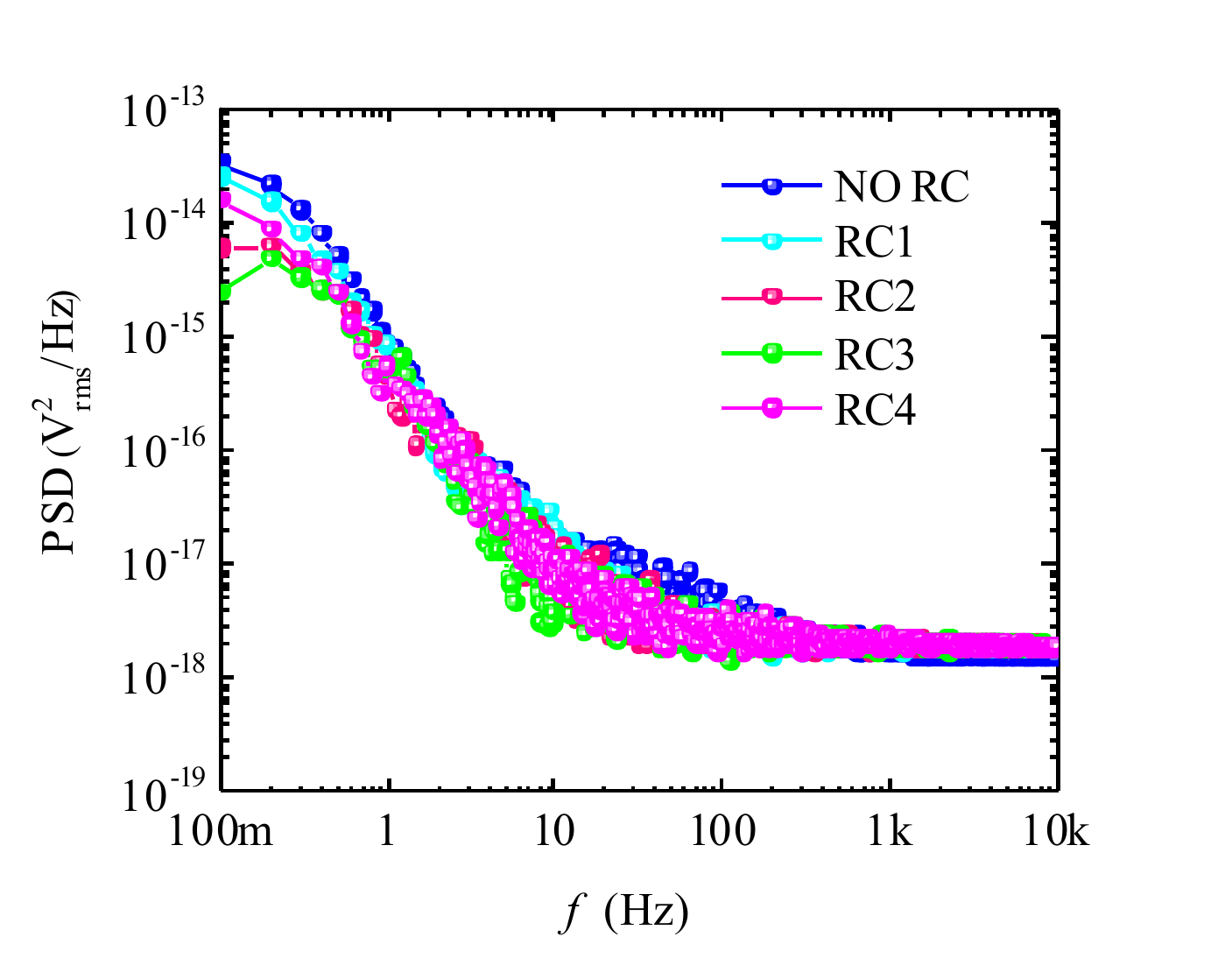}
\end{center}
  \caption{Cross-spectra for a short circuit measured for $60\,$min with different high-pass RC filter settings (see Table~\ref{tab:RC}. So long as the cutoff frequency $f_c$ is well below the frequency of interest, the spectra are nominally identical). \label{fig:RC} }
\end{figure}

In principle, one should also consider the contribution from the high-pass filter in the preamplifiers, which would serve to roll off the spectrum at lower frequencies. In practice, the high-pass filter is set at least 1 decade below the lowest frequency of interest, which is itself determined by the buffering time, sufficient to make the filter's contribution negligible. In figure~\ref{fig:RC} we show the effect of different RC filter settings (see Table~\ref{tab:RC} for details on the filter settings) on the measured spectra, and indeed so long as the filter cutoff frequency $f_c$ is well below the frequency of interest, the high-pass filter has a negligible effect. 

\begin{table}[h]
	\caption{RC high-pass filter settings for the EG\&G 113 preamplifiers}
	\begin{ruledtabular}
	\begin{tabular}{cccc}
	\textrm{\bf Setting} & \textrm{\bf R (M$\Omega$)} & \textrm{\bf C ($\mu$F)} & \textrm{\bf $f_c$ (mHz)} \\
	\colrule
	\rule{0pt}{2.6ex} RC1 & 1.50 & 3.30 & 32.2 \\
	\rule{0pt}{2.6ex} RC2 & 1.50 & 10.0 & 10.6 \\
	\rule{0pt}{2.6ex} RC3 & 15.9 & 3.30 & 3.03\\
	\rule{0pt}{2.6ex} RC4 & 15.9 & 10.0 & 1.00\\
	\end{tabular} \label{tab:RC}
	\end{ruledtabular}
\end{table}

\begin{figure}[h!]
\begin{center}
  \includegraphics[width=0.95\columnwidth]{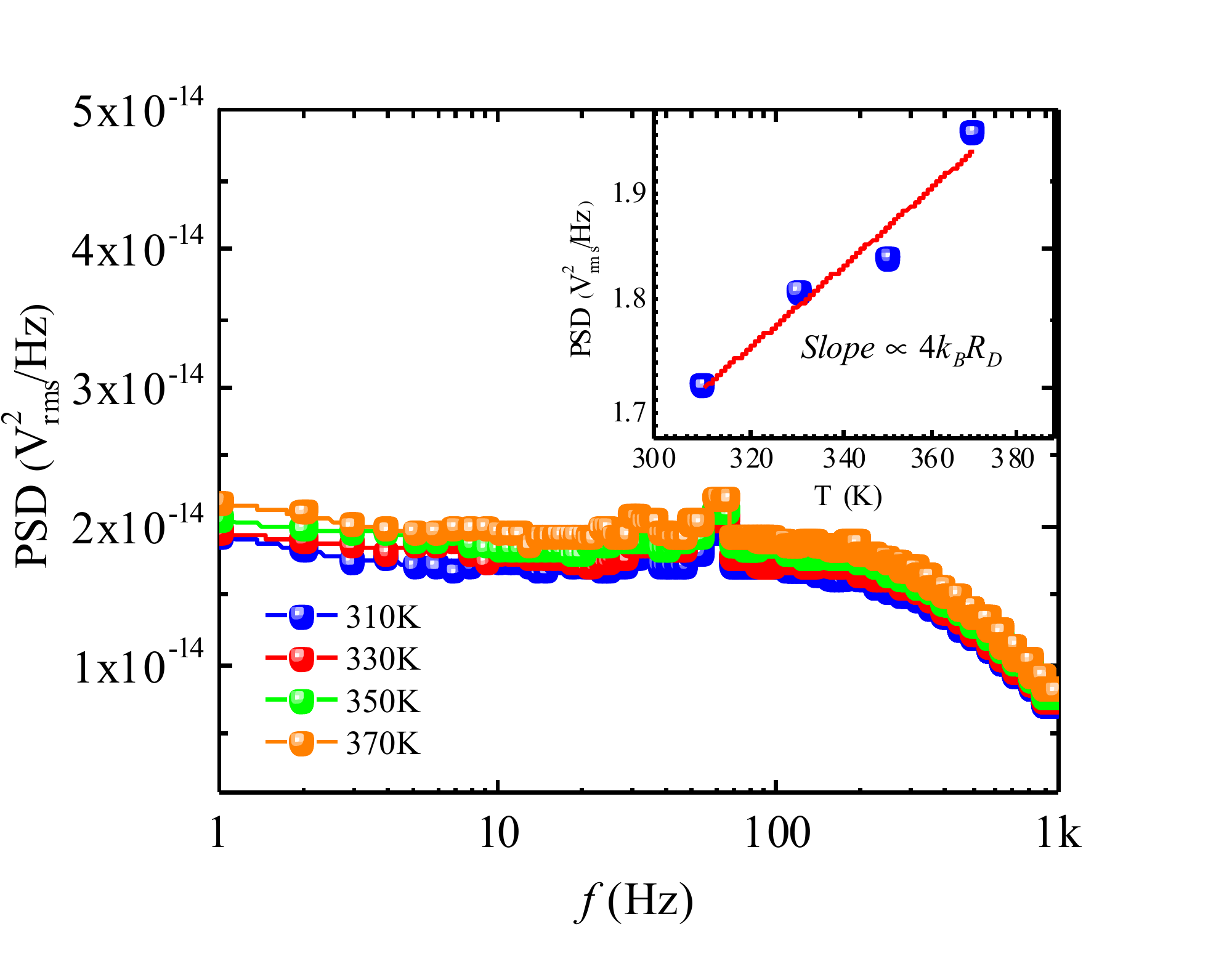}
\end{center}
  \caption{Power spectral density for a $1\,\text{M}\Omega$ resistor measured for $10\,$min for various temperatures. Inset: magnitude of the average power spectral density over the low frequency range versus temperature. The PSD scales linearly with temperature, consistent with Johnson-Nyquit theorem, and the slope is given by $4k_BR$.}\label{fig:t-dep}
\end{figure}

The results above demonstrate that the power spectral density indeed scales with resistance as expected, within the limits of instrument sensitivity and the frequency limits imposed by the preamplifiers and the device itself. According  to Johson-Nyquist theorem, temperature can also affect the noise level (Fig.~\ref{fig:t-dep}). In order to study the effect of temperature on the noise, specifically for devices with temperature-dependent impedance,  we designed a sample holder with temperature control using a resistive heater (a $50\,$W, $25\,\Omega$ chassis mount resistor) and a diode temperature sensor (Lake Shore DT-600 series) which can set the temperature of the sample using our software with relative accuracy of $0.1\,$K and stability of $0.01\,$K. More information on the temperature control is provided in Appendix \ref{sec:temperature}. Figure~\ref{fig:t-dep} shows the power spectral density measured for $10\,$min for a $1\,\text{M}\Omega$ carbon film resistor at various temperatures. We determined the white noise contribution to the PSD by finding the average of the PSD over the lower frequency ($<\!100\,$Hz) portion of the spectrum, where the roll-off from the low pass filter is negligible. Though the temperature range is limited ($310\!-\!370\,$K), the PSD does scale approximately linearly with temperature, as shown in the inset to Fig.~\ref{fig:t-dep}. 

In this case, the thermal noise is well above the sensitivity limit of the instrument, and little averaging is required. In addition, focusing only on the thermal noise one does not need to measure to very low frequencies, a $1\,$Hz minimum frequency will suffice, meaning that the buffering times may be reduced as well. Overall, the effect is that suitable spectra may be obtained in only $10\,$min. 

\section{Noise Components Analysis}\label{sec:1f}

In addition to the Johnson-Nyquist frequency-independent or ``white" noise, excess low frequency noise is typically observed when there is current in the device under test. This noise typically follows a $S_v\!\propto\!1/f$ spectrum, hence its name of ``$1/f$ noise, also known as flicker noise.\cite{Hooge1981,Dutta1981,McWhorter1955}, Unlike Johnson-Nyquist noise, this contribution is frequency dependent and more apparent in the low frequency regimes. Because of the presence and ubiquity of this noise in many systems, either man-made or in nature, a great deal of research has been done to search for a universal explanation for the origin of this noise, and as yet no common origin has been found.\cite{Dutta1981} Another baffling feature of this noise is its divergence at low frequencies and the apparent absence of a low frequency cut off. The most well-known form for parameterizing $1/f$ noise is HoogeÕs phenomenological formula, 

\begin{equation}
S_v = \frac{V^{2+\beta}}{N_C f^\alpha}
\end{equation}

Where $V$ is the voltage on the device under test, $N_C$ the carrier concentration, and $\alpha$ and $\beta$ are constants, with $\alpha\!\sim\!1$ and $\beta\!\sim\!0$.\cite{Hooge1981,Dutta1981} While the origin of this noise contribution is often mysterious, it may still serve as a useful probe for microscopic studies of a system.\cite{Akiba1997,Kleinpenning1985,Soliveres2007,Takagi1986,McWhorter1955,Fleetwood1984}

In order to measure the $1/\freq$ noise of a device resulting from a dc current, it is a common design to bias the device with a current source.\cite{Agilent} Notably, the source output impedance should be much higher than the device as discussed below. To obtain the best performance for preamplifiers, the dc offset at testing port can be either balanced by current compensation or by a passive high-pass filter. In this design, we chose the latter for convenience. A more effective way to cancel the dc offset is by employing a Wheatstone bridge\cite{Jonker1999}, however this design requires identical or carefully patterned samples.

\begin{figure}[h]
\begin{center}
  \includegraphics[width=0.95\columnwidth]{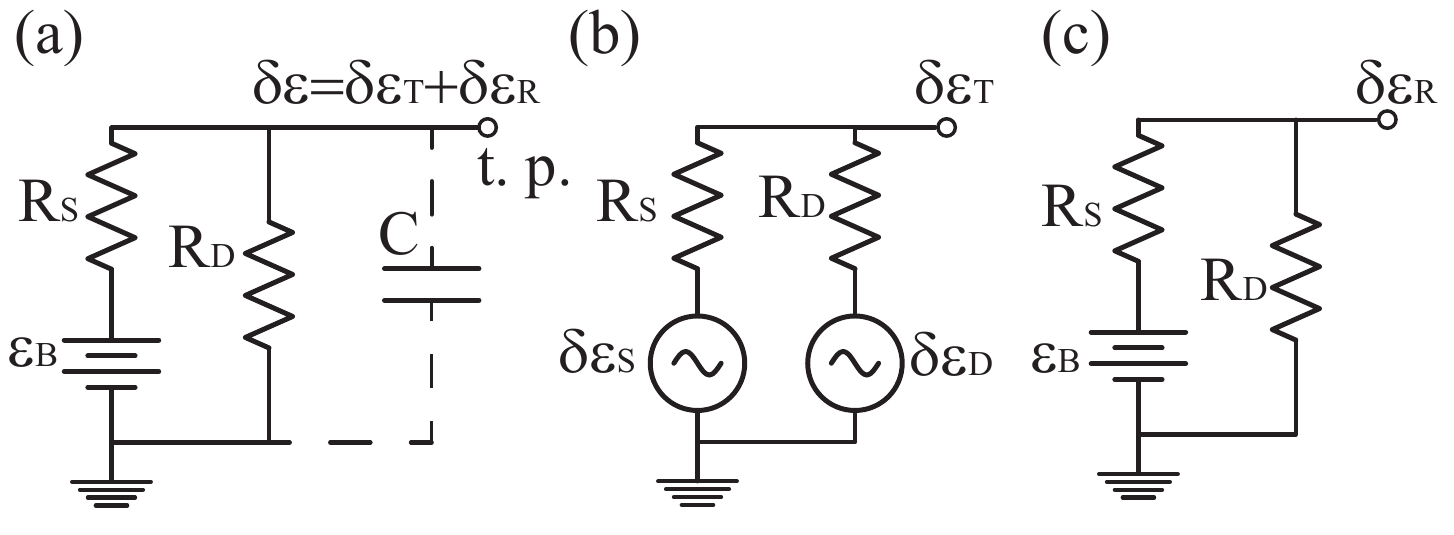}  
\end{center} 
  \caption{Equivalent circuit of  {(a)}$\,$measurement setup with current source constructed from the battery $\varepsilon_B$ and series resistor $R_S$. The total noise has contributions from thermal noise and resistance fluctuations.  {(b)} Total thermal noise contribution $\delta \varepsilon_T$ arising from equivalent thermal noise generators $\delta \varepsilon_S$ and $\delta \varepsilon_R$ in $R_S$ and $R_D$.  {(c)} The additional noise contribution $\delta \varepsilon_R$ arising from resistance fluctuations in $R_S$ and $R_D$. t.p. = testing port.\label{fig:eq-circuit}}
\end{figure}

The equivalent circuit of this analyzer with a high impedance, battery-driven current source is described in Fig.~\ref{fig:eq-circuit}(a). $R_D$ and $R_S$ are the device and source resistances, respectively, $\e_B$ is the battery voltage, and $C$ the stray capacitance present between the testing port and ground. The voltage fluctuation$\,$($\de$) at the testing port can be seen as the superposition of thermal noise$\,$($\de_T$) and noise arising from fluctuation of resistance$\,$($\de_R$). Assuming $\de_T$ and $\de_R$ are orthogonal$\,$($\Mean{\de_T\cdot\de_R}=0$), the mean square of $\de$ is the sum of $\Mean{\de_T^2}$ and $\Mean{\de_R^2}$:

\begin{equation}
\MeanSquare{\de} = \MeanSquare{\de_T} + \MeanSquare{\de_R}
\label{eqn:DecompositeVoltageNoise}
\end{equation}

The thermal noise is a fluctuation-dissipation phenomenon, whose power spectral density level \emph{only} depends on the equivalent resistance of the circuit.\cite{Nyquist1928} In our case, as shown in Fig.~\ref{fig:eq-circuit}(b), the resistance measured between the ground and the testing port of this circuit is equal to the parallel resistance of device and the source resistor ($R_P\!=\!R_S R_D/(R_S+R_D)$), thus the apparent thermal noise level $\de_{T}$ is:

\begin{equation}
\MeanSquare{ \de_T }  = {4k_BTR_P}
\label{eqn:deTMeanSquare}
\end{equation}

If $R_S{\gg}R_D$, $R_P{\approx}R_D$, the thermal noise at testing port will be close to that considered generated by the device.

Figure \ref{fig:eq-circuit}(c) shows the equivalent circuit to analysis the noise arising from fluctuation of resistance.
The steady-state equilibrium voltage at the output terminal is $\varepsilon\!=\!\varepsilon_B R_D/(R_D+R_S)$. We denote the resistance fluctuations in $R_S$ and $R_D$ be $\delta R_S$ and $\delta R_D$, respectively, and assume that the battery voltage $\varepsilon_B$ is constant ($\delta \varepsilon_B\!=\!0$). 
Presuming the fluctuations in $R_S$ and $R_D$ to be independent, the noise arising from resistance fluctuations is:

\begin{equation}
\begin{aligned}
\MeanSquare{\de_R} =& \left(\frac{\partial \varepsilon}{\partial R_S}\right)^2 \,\MeanSquare{\delta R_S}  + \left(\frac{\partial \varepsilon}{\partial R_D} \right)^2 \, \MeanSquare{\delta R_D}\hfill\\
                   =& \frac{\varepsilon_B^2}{\left(R_S+R_D\right)^2} \biggl[\frac{R_D^2}{\left(R_S+R_D\right)^2}\MeanSquare{\delta R_S} + ... \hfill\\
                   &... +\frac{R_S^2 \, }{\left(R_S+R_D\right)^2} \MeanSquare{\delta R_D}\biggr]
\end{aligned}
\label{eqn:de_dR}
\end{equation}

It is more convenient to express the results in terms of the relative resistance fluctuations $x_S\!\equiv\!\delta R_S/R_S$ and $x_D\!\equiv\!\delta R_D/R_D$. Noting also that the steady-state dc current $I$ through $R_S$ and $R_D$ is $I\!=\!\varepsilon_B/(R_S+R_D)$ and again $R_P\!=\!R_S R_D/(R_S+R_D)$, Eqn.~\ref{eqn:de_dR} simplifies to

\begin{equation} 
\MeanSquare{\de_R} = I^2R_P^2 \left(\MeanSquare{\x_S} + \MeanSquare{\x_D}\right)
\label{eqn:deRMeanSquare}
\end{equation}

In a simplest scenario, it is assumed the instantaneous resistance fluctuations are proportional to the resistance itself, but remain functions of frequency ($\delta R/R = \x(f)$), and the function $\x$ then characterizes the frequency dependence of instantaneous resistance fluctuation ratio ($\delta R/R$). 

This reproduces a well-known experimental result, namely that the mean square of voltage fluctuation caused by resistance fluctuation revealed by passing current ($\de_{R}^2$) is proportional to $I^2$ and $R^{2}$.\cite{Hooge1981} However, it should be stressed that as far as this equation holds, $\x_D$ can be measured only when it is much greater than $\x_S$ or when $\x_S$ is already known. Substituting Eqn.~\ref{eqn:deTMeanSquare} and \ref{eqn:deRMeanSquare} into Eqn.~\ref{eqn:DecompositeVoltageNoise}, and including the effect of the stray capacitance $C$ simply multiplies the entire response by the effective low-pass filter response (Eqn.~\ref{eqn:SpectrumShapedByLPF}), the total noise at testing port is given by:

\begin{equation}
\MeanSquare{\de} = \frac{{4{k_B}T{R_P} + {I^2}R_P^2(\MeanSquare{\x_D} + \MeanSquare{\x_S})}}{{1 + {{(2\pi \freq {R_P}C)}^2}}}
\label{eqn:deMeanSquare}
\end{equation}

In an attempt to get an understanding of $\MeanSquare{\x_S}$, cross-spectrum is measured when $R_D$ is substituted by a resistor identical to $R_S$. In this arrangement, the mean square value of resistance fluctuation ratio of both resistors should be the same, however, their instantaneous fluctuations are still independent of each other. It is easy to deduce from Eqn.~\ref{eqn:deMeanSquare} that if $R_D$ is the same as $R_S$, and further assuming $\MeanSquare{\x_S}$ has a current-independent $1/\freq$ characteristic$\,$($\MeanSquare{\x_S}  = {k_S}{\freq ^{ - \alpha }}$), the noise voltage signal at testing port is:

\begin{equation}
\MeanSquare{\de} = \frac{{4{k_B}T{R_S} + {I^2}R_S^2{k_S}{f^{ - \alpha }}}}{{2 + {{(\sqrt{2}\pi \freq {R_S}C)}^2}}}
\label{eqn:deRs}
\end{equation}

\begin{figure}[h]
\begin{center}
  \includegraphics[width=0.95\columnwidth]{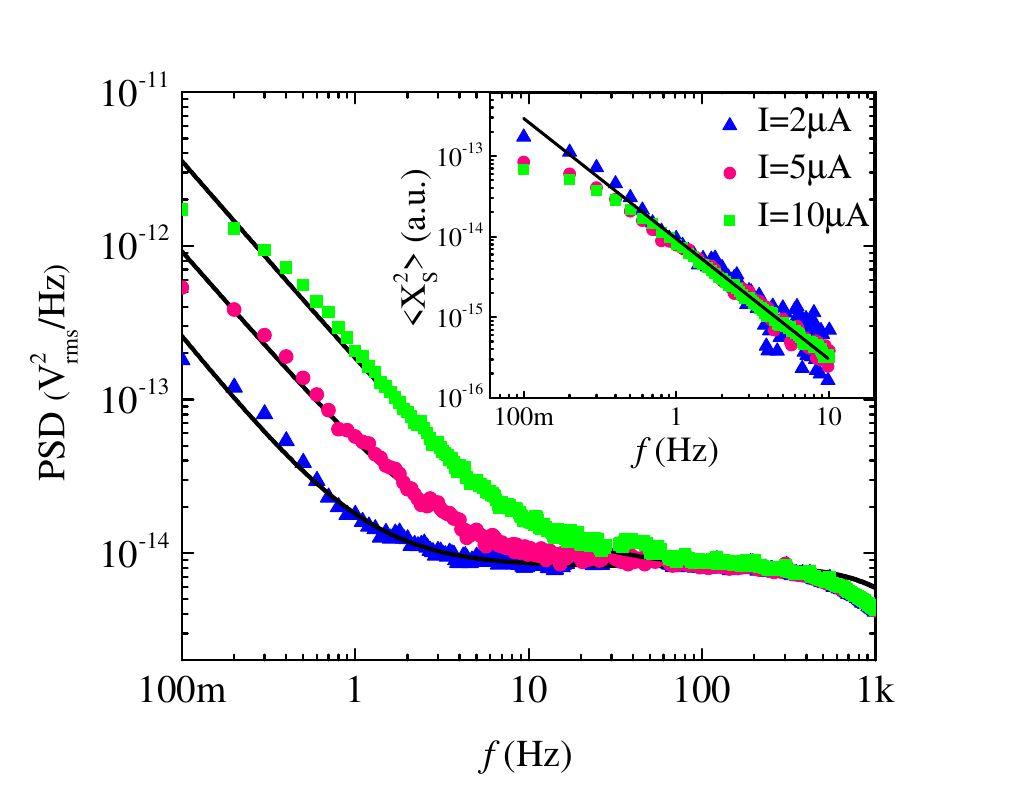}
\end{center}
  \caption{ Measured noise spectra when $R_D$ is identical to $R_S$(\SI{1}{\mega\Ohm} metal film resistor) at different applied current settings.  {Insert:} Extracted frequency dependence of resistance fluctuation ratio of source resistor from Eqn.~\ref{eqn:deRs}.\label{fig:source-noise} }
\end{figure}

Figure \ref{fig:source-noise} shows measured noise spectra when both of $R_D$ and $R_S$ are identical \SI{1}{\mega\Ohm} metal film resistors (Vishay RNC55\cite{VishayDale}) for three different dc current levels. The data showed an excellent fit to Eqn.~\ref{eqn:deRs} with $k_s\!=\!9.42({\pm}0.01){\times}10^{-15}$ and $\alpha\!=1.49({\pm}0.04)$. The resistance fluctuation of this kind of resistors is verified to be very low$\,$(about $0.04\,$ppm at $1\,Hz$), assuring they are ideal for low level noise testing purposes. The extracted $\MeanSquare{\x_S}$ of different current settings (Inset of Fig.~\ref{fig:source-noise}) are basically consistent, proving the assumptions we made about $\MeanSquare{\x_S}$ are reasonable.

\begin{figure}[h]
\begin{center}
  \includegraphics[width=0.95\columnwidth]{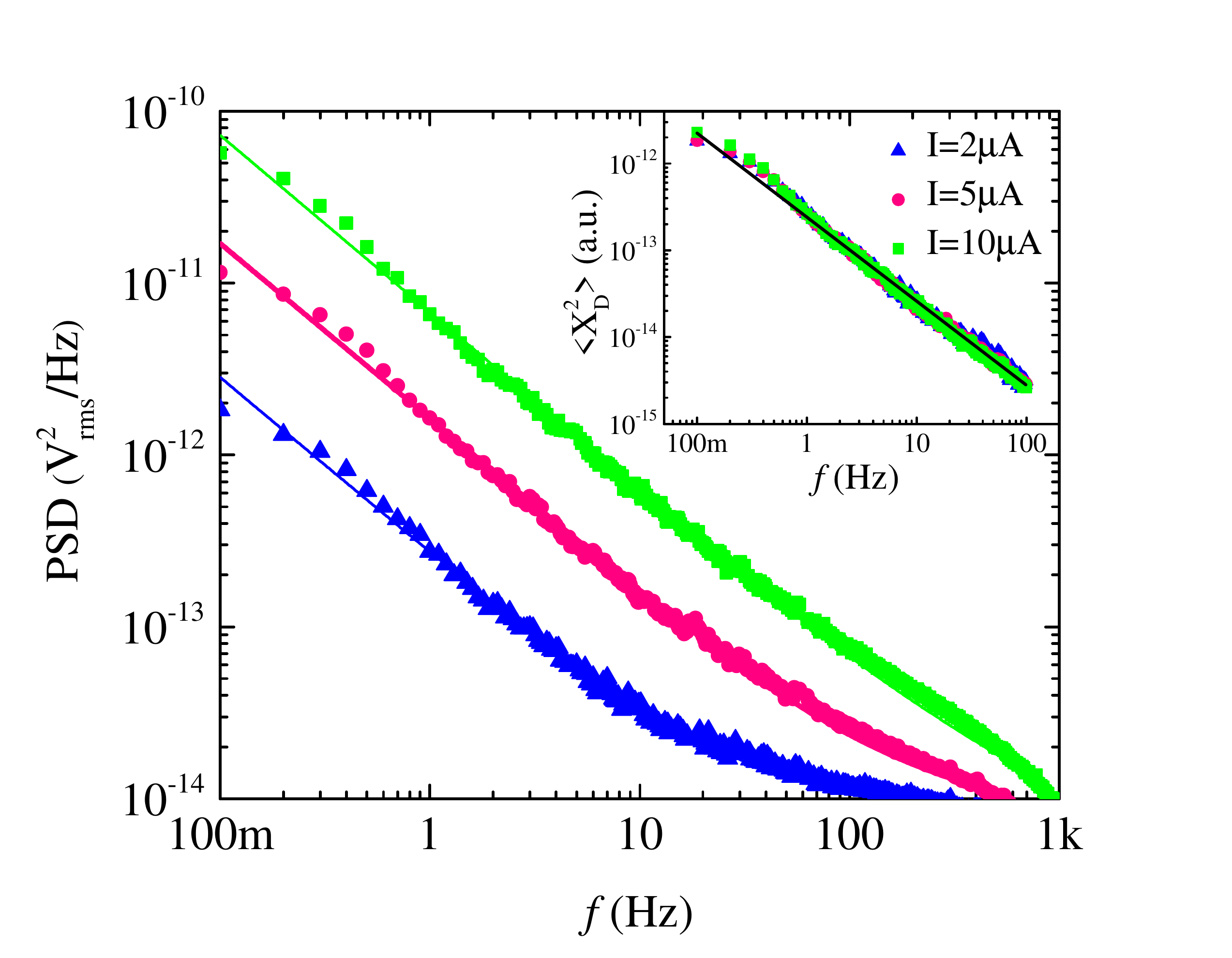}
\end{center}
  \caption{ Measured noise spectra of a (\SI{1}{\mega\Ohm} carbon film resistor) at different applied current settings. {Insert:} Extracted frequency dependence of resistance fluctuation ratio from Eqn.~\ref{eqn:deMeanSquare}.
  \label{fig:noisy-resistor}}
\end{figure}

In comparison with carbon film resistors used in the current source, we also measured noise spectra of a standard $5\$, \SI{1}{\mega\Ohm}$ carbon film resistors for various dc current levels (Fig.~\ref{fig:noisy-resistor}). Despite that they have the same rated resistances, the extracted $k_s$ for this resistor is $2.41(\pm 0.09)\times 10^{-13}$, more than $250$ times higher than the Vishay RNC55. The exponential component is found to be slightly lower, and closer to the usual ``$1/f$'' value, viz. $\alpha\!=\!0.97(\pm 0.01)$).

\section{Conclusion}
We report on the design and implementation of a simple, flexible, and highly sensitive cross-spectrum analyzer. 
The time dependence of sensitivity limit for this system is carefully characterized to perform reliable measurements for low frequency noise spectra. Due to the frequency-dependent characteristic of the background noise, it is of critical important to rule out this contribution from the measured spectra.
A systematic routine for noise components analysis is provided and the noise spectra of current source and carbon resistors are examined with this method and a $1/f$ resistivity fluctuation with exponential components $\sim\!1$ are identified.

\appendix 

\section{Measurement environment, shielding, and grounding} \label{sec:shielding}

\begin{figure}[h]
\begin{center}
  \includegraphics[width=0.95\columnwidth]{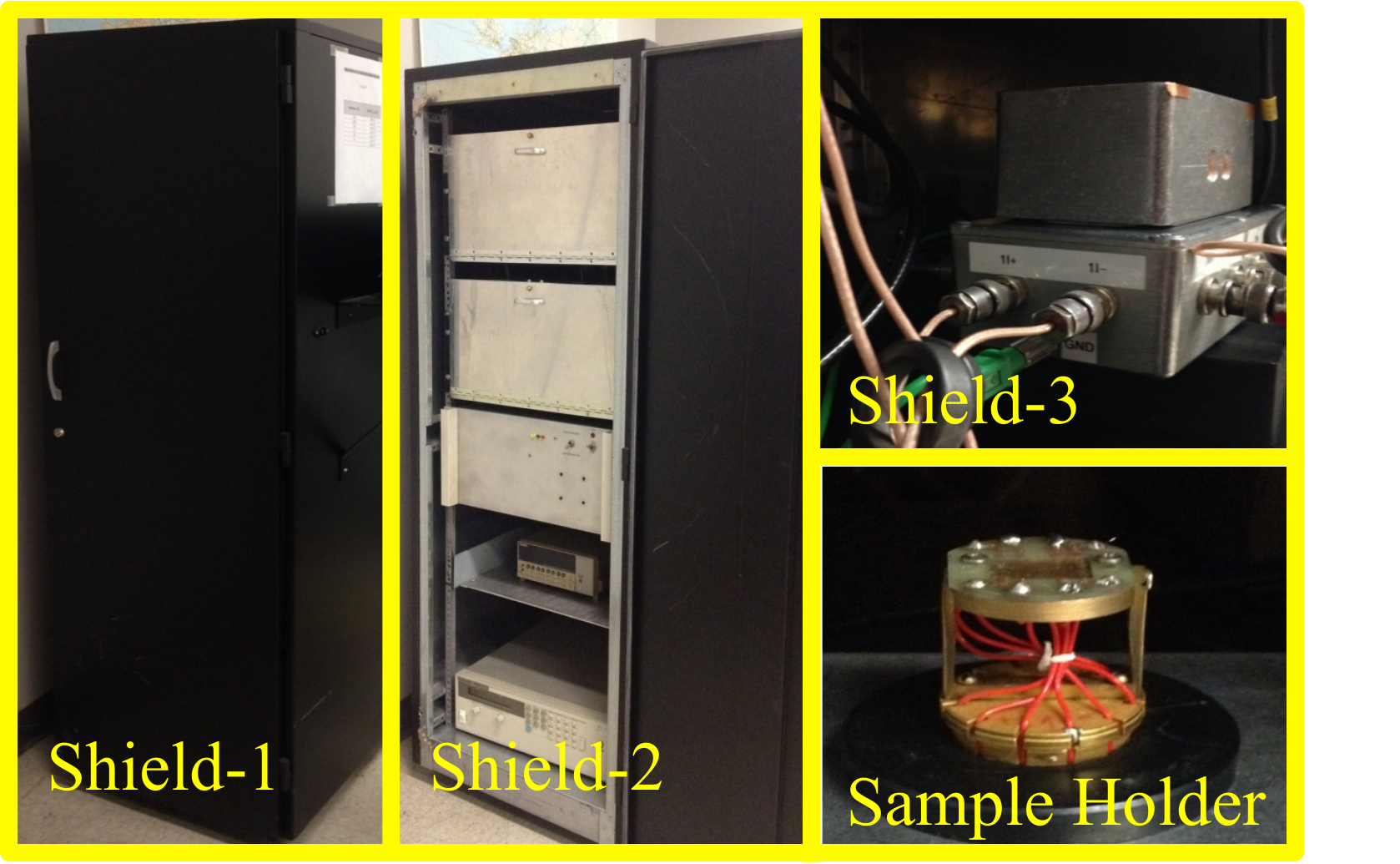}
\end{center}
  \caption{Shielding the instrument and the device under the test from environmental electromagnetic interference. Two layers of shielding, shield 1 and 2 contain the all components of the analyzer, and the third layer of shielding around the sample holder further minimizes  external influences. 
}\label{fig:setup}
\end{figure}

Canceling the spurious environmental noise, through shielding and grounding is an important step in noise measurement of devices. As shown in Fig.\ref{fig:setup},  our noise spectrum analyzer is shielded within two conducting boxes to shield the whole instrument from surrounding electromagnetic interference. The first outer box is a ferrous steel, which serves to eliminate electric field effects and minimize magnetic induction effects. The second inner boxes (one for the sample, and a second for the preamplifiers and DAQ) are aluminum, and serve to further decouple the sample and amplifiers from the outside as well as minimizing crosstalk between them. The device under the test is further shielded by a third laminated box, with outer ferrous steel and inner copper shields, to further minimize the environmental effects on the measurements. In order to avoid power-line voltage fluctuations coming from the power outlets or $60\,$Hz multipliers induced by electromagnetic waves from the current passing through the power cables, all the components in our analyzer are disconnected from the power outlet and powered by batteries, with sufficient capacity to perform measurements of up to 24 hours. Remaining components requiring an ac power connection (e.g., computer, heater power supply) are powered through an isolation transformer which is itself heavily shielded from the box containing the analyzer itself. All the components are properly grounded to avoid the buildup of the excess charge and minimize ground loop noise, and maximum consideration is implemented for reducing the environmental noise as much as possible. Proper choice of location for the experiment is another factor affecting the measurements, i.e. it should be far from any significant source of electromagnetic radiation that could affect the measurement results substantially. 

\section{Sample temperature control} \label{sec:temperature}

Our temperature control system is relatively straightforward, the primary considerations being (1) high stability of temperature once a set point has been reached ($<\!0.01\,$K) to avoid introducing spurious thermal effects, and (2) high relative accuracy of the temperature set points ($<\!0.1\,$K). Components of our temperature-control system include: a heating unit made of a resistive heater element (a $50\,$W, $25\,\Omega$ chassis mount resistor) along with a DC power supply (HP-6654A) as the power source; a diode temperature sensor (Lakeshore DT-600 series) with high sensitivity of $0.01\,$K; and a customized PID temperature control loop written in LabView to adjust the sample temperature with relative accuracy of $\pm\!0.1\,$K for temperature changes in the range of $300\!-\!400\,$K.

\begin{figure}[h]
\begin{center}
  \includegraphics[width=0.95\columnwidth]{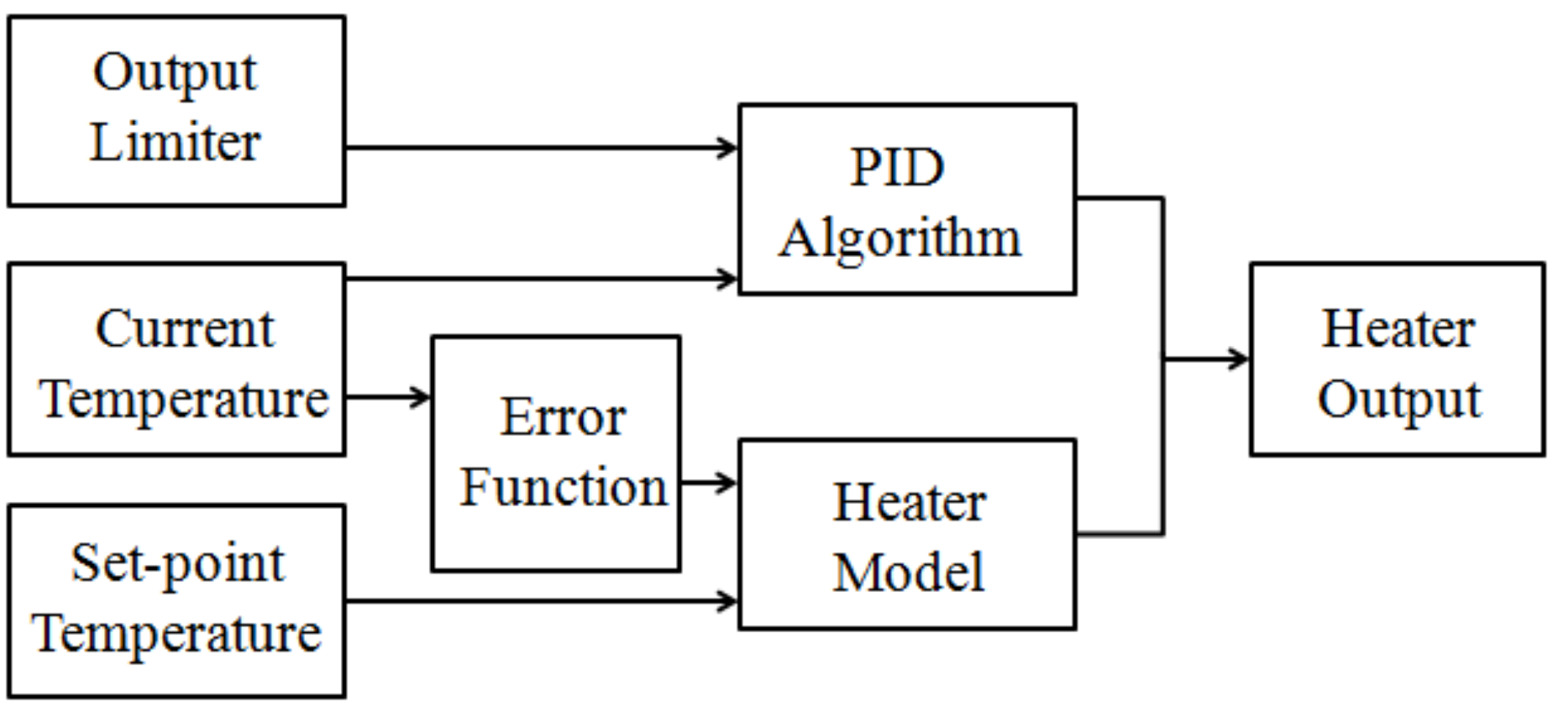}
\end{center}
  \caption{Schematic of the temperature control procedure.}\label{fig:t-con}
\end{figure}

Using our  temperature control LabView VI, heating and cooling of the samples is possible through different approaches, such as ``Fast settle'' and ``No-overshoot.'' These approaches can be selected depending on sensitivity of the device under the test to temperature overshoot, or out of necessity to limit the settling time. These requirements were achieved through the addition of a heater model to the PID algorithm.  The heater model was created by recording the steady state output of our power supply as a function of the set point temperature. This model is used to bias the power output to the expected value for a given temperature. Overshoot is further limited by controlling the maximum power adjustment to our heater model by the PID controller, generally set to $\pm\!10\,\%$ for minimal overshoot. This value can be increased to enhance settling speed for samples where overshoot is not an important consideration. Figure~\ref{fig:t-con} shows a schematic of the temperature control procedure.

\section*{Acknowledgment}
The authors gratefully acknowledge D.\ Whitcomb for his help in building the system.

\bibliography{refs}

\begin{thebibliography}{25}%
\makeatletter
\providecommand \@ifxundefined [1]{%
 \@ifx{#1\undefined}
}%
\providecommand \@ifnum [1]{%
 \ifnum #1\expandafter \@firstoftwo
 \else \expandafter \@secondoftwo
 \fi
}%
\providecommand \@ifx [1]{%
 \ifx #1\expandafter \@firstoftwo
 \else \expandafter \@secondoftwo
 \fi
}%
\providecommand \natexlab [1]{#1}%
\providecommand \enquote  [1]{``#1''}%
\providecommand \bibnamefont  [1]{#1}%
\providecommand \bibfnamefont [1]{#1}%
\providecommand \citenamefont [1]{#1}%
\providecommand \href@noop [0]{\@secondoftwo}%
\providecommand \href [0]{\begingroup \@sanitize@url \@href}%
\providecommand \@href[1]{\@@startlink{#1}\@@href}%
\providecommand \@@href[1]{\endgroup#1\@@endlink}%
\providecommand \@sanitize@url [0]{\catcode `\\12\catcode `\$12\catcode
  `\&12\catcode `\#12\catcode `\^12\catcode `\_12\catcode `\%12\relax}%
\providecommand \@@startlink[1]{}%
\providecommand \@@endlink[0]{}%
\providecommand \url  [0]{\begingroup\@sanitize@url \@url }%
\providecommand \@url [1]{\endgroup\@href {#1}{\urlprefix }}%
\providecommand \urlprefix  [0]{URL }%
\providecommand \Eprint [0]{\href }%
\providecommand \doibase [0]{http://dx.doi.org/}%
\providecommand \selectlanguage [0]{\@gobble}%
\providecommand \bibinfo  [0]{\@secondoftwo}%
\providecommand \bibfield  [0]{\@secondoftwo}%
\providecommand \translation [1]{[#1]}%
\providecommand \BibitemOpen [0]{}%
\providecommand \bibitemStop [0]{}%
\providecommand \bibitemNoStop [0]{.\EOS\space}%
\providecommand \EOS [0]{\spacefactor3000\relax}%
\providecommand \BibitemShut  [1]{\csname bibitem#1\endcsname}%
\let\auto@bib@innerbib\@empty
%</preamble>
\bibitem [{\citenamefont {Hooge}\ \emph {et~al.}(1981)\citenamefont {Hooge},
  \citenamefont {Kleinpenning},\ and\ \citenamefont {Vandamme}}]{Hooge1981}%
  \BibitemOpen
  \bibfield  {author} {\bibinfo {author} {\bibfnamefont {F.~N.}\ \bibnamefont
  {Hooge}}, \bibinfo {author} {\bibfnamefont {T.~G.~M.}\ \bibnamefont
  {Kleinpenning}}, \ and\ \bibinfo {author} {\bibfnamefont {L.~K.~J.}\
  \bibnamefont {Vandamme}},\ }\href
  {http://stacks.iop.org/0034-4885/44/i=5/a=001} {\bibfield  {journal}
  {\bibinfo  {journal} {Reports on Progress in Physics}\ }\textbf {\bibinfo
  {volume} {44}},\ \bibinfo {pages} {479} (\bibinfo {year} {1981})}\BibitemShut
  {NoStop}%
\bibitem [{\citenamefont {Bess}(1956)}]{Bess1956}%
  \BibitemOpen
  \bibfield  {author} {\bibinfo {author} {\bibfnamefont {L.}~\bibnamefont
  {Bess}},\ }\href {\doibase 10.1103/PhysRev.103.72} {\bibfield  {journal}
  {\bibinfo  {journal} {Phys. Rev.}\ }\textbf {\bibinfo {volume} {103}},\
  \bibinfo {pages} {72} (\bibinfo {year} {1956})}\BibitemShut {NoStop}%
\bibitem [{\citenamefont {Mihaila}(2000)}]{Mihaila2000}%
  \BibitemOpen
  \bibfield  {author} {\bibinfo {author} {\bibfnamefont {M.}~\bibnamefont
  {Mihaila}},\ }in\ \href {\doibase 10.1007/3-540-45463-2_11} {\emph {\bibinfo
  {booktitle} {Noise, Oscillators and Algebraic Randomness}}},\ \bibinfo
  {series} {Lecture Notes in Physics}, Vol.\ \bibinfo {volume} {550},\ \bibinfo
  {editor} {edited by\ \bibinfo {editor} {\bibfnamefont {M.}~\bibnamefont
  {Planat}}}\ (\bibinfo  {publisher} {Springer Berlin Heidelberg},\ \bibinfo
  {year} {2000})\ pp.\ \bibinfo {pages} {216--231}\BibitemShut {NoStop}%
\bibitem [{\citenamefont {Mazumdar}\ \emph {et~al.}(2007)\citenamefont
  {Mazumdar}, \citenamefont {Liu}, \citenamefont {Schrag}, \citenamefont
  {Carter}, \citenamefont {Shen},\ and\ \citenamefont {Xiao}}]{Mazumdar2007}%
  \BibitemOpen
  \bibfield  {author} {\bibinfo {author} {\bibfnamefont {D.}~\bibnamefont
  {Mazumdar}}, \bibinfo {author} {\bibfnamefont {X.}~\bibnamefont {Liu}},
  \bibinfo {author} {\bibfnamefont {B.~D.}\ \bibnamefont {Schrag}}, \bibinfo
  {author} {\bibfnamefont {M.}~\bibnamefont {Carter}}, \bibinfo {author}
  {\bibfnamefont {W.}~\bibnamefont {Shen}}, \ and\ \bibinfo {author}
  {\bibfnamefont {G.}~\bibnamefont {Xiao}},\ }\href {\doibase
  10.1063/1.2754352} {\bibfield  {journal} {\bibinfo  {journal} {Applied
  Physics Letters}\ }\textbf {\bibinfo {volume} {91}},\ \bibinfo {eid} {033507}
  (\bibinfo {year} {2007})}\BibitemShut {NoStop}%
\bibitem [{\citenamefont {Guerrero}\ \emph {et~al.}(2006)\citenamefont
  {Guerrero}, \citenamefont {Aliev}, \citenamefont {Tserkovnyak}, \citenamefont
  {Santos},\ and\ \citenamefont {Moodera}}]{Guerrero2006}%
  \BibitemOpen
  \bibfield  {author} {\bibinfo {author} {\bibfnamefont {R.}~\bibnamefont
  {Guerrero}}, \bibinfo {author} {\bibfnamefont {F.~G.}\ \bibnamefont {Aliev}},
  \bibinfo {author} {\bibfnamefont {Y.}~\bibnamefont {Tserkovnyak}}, \bibinfo
  {author} {\bibfnamefont {T.~S.}\ \bibnamefont {Santos}}, \ and\ \bibinfo
  {author} {\bibfnamefont {J.~S.}\ \bibnamefont {Moodera}},\ }\href
  {http://link.aps.org/doi/10.1103/PhysRevLett.97.266602} {\bibfield  {journal}
  {\bibinfo  {journal} {Phys. Rev. Lett.}\ }\textbf {\bibinfo {volume} {97}},\
  \bibinfo {pages} {266602} (\bibinfo {year} {2006})}\BibitemShut {NoStop}%
\bibitem [{\citenamefont {Kleinpenning}(1985)}]{Kleinpenning1985}%
  \BibitemOpen
  \bibfield  {author} {\bibinfo {author} {\bibfnamefont {T.~G.~M.}\
  \bibnamefont {Kleinpenning}},\ }\href {\doibase 10.1116/1.573194} {\bibfield
  {journal} {\bibinfo  {journal} {J. Vac. Sci. Technol. A}\ }\textbf {\bibinfo
  {volume} {3}},\ \bibinfo {pages} {176} (\bibinfo {year} {1985})}\BibitemShut
  {NoStop}%
\bibitem [{\citenamefont {Akiba}(1997)}]{Akiba1997}%
  \BibitemOpen
  \bibfield  {author} {\bibinfo {author} {\bibfnamefont {M.}~\bibnamefont
  {Akiba}},\ }\href {\doibase 10.1063/1.120301} {\bibfield  {journal} {\bibinfo
   {journal} {Applied Physics Letters}\ }\textbf {\bibinfo {volume} {71}},\
  \bibinfo {pages} {3236} (\bibinfo {year} {1997})}\BibitemShut {NoStop}%
\bibitem [{\citenamefont {Rammal}\ \emph {et~al.}(1985)\citenamefont {Rammal},
  \citenamefont {Tannous}, \citenamefont {Breton},\ and\ \citenamefont
  {Tremblay}}]{Rammal1985}%
  \BibitemOpen
  \bibfield  {author} {\bibinfo {author} {\bibfnamefont {R.}~\bibnamefont
  {Rammal}}, \bibinfo {author} {\bibfnamefont {C.}~\bibnamefont {Tannous}},
  \bibinfo {author} {\bibfnamefont {P.}~\bibnamefont {Breton}}, \ and\ \bibinfo
  {author} {\bibfnamefont {A.~M.~S.}\ \bibnamefont {Tremblay}},\ }\href
  {\doibase 10.1103/PhysRevLett.54.1718} {\bibfield  {journal} {\bibinfo
  {journal} {Phys. Rev. Lett.}\ }\textbf {\bibinfo {volume} {54}},\ \bibinfo
  {pages} {1718} (\bibinfo {year} {1985})}\BibitemShut {NoStop}%
\bibitem [{\citenamefont {Soliveres}\ \emph {et~al.}(2007)\citenamefont
  {Soliveres}, \citenamefont {Gyani}, \citenamefont {Delseny}, \citenamefont
  {Hoffmann},\ and\ \citenamefont {Pascal}}]{Soliveres2007}%
  \BibitemOpen
  \bibfield  {author} {\bibinfo {author} {\bibfnamefont {S.}~\bibnamefont
  {Soliveres}}, \bibinfo {author} {\bibfnamefont {J.}~\bibnamefont {Gyani}},
  \bibinfo {author} {\bibfnamefont {C.}~\bibnamefont {Delseny}}, \bibinfo
  {author} {\bibfnamefont {A.}~\bibnamefont {Hoffmann}}, \ and\ \bibinfo
  {author} {\bibfnamefont {F.}~\bibnamefont {Pascal}},\ }\href {\doibase
  10.1063/1.2709853} {\bibfield  {journal} {\bibinfo  {journal} {Applied
  Physics Letters}\ }\textbf {\bibinfo {volume} {90}},\ \bibinfo {eid} {082107}
  (\bibinfo {year} {2007})}\BibitemShut {NoStop}%
\bibitem [{\citenamefont {Podzorov}\ \emph {et~al.}(2001)\citenamefont
  {Podzorov}, \citenamefont {Gershenson}, \citenamefont {Uehara},\ and\
  \citenamefont {Cheong}}]{Podzorov2001}%
  \BibitemOpen
  \bibfield  {author} {\bibinfo {author} {\bibfnamefont {V.}~\bibnamefont
  {Podzorov}}, \bibinfo {author} {\bibfnamefont {M.~E.}\ \bibnamefont
  {Gershenson}}, \bibinfo {author} {\bibfnamefont {M.}~\bibnamefont {Uehara}},
  \ and\ \bibinfo {author} {\bibfnamefont {S.-W.}\ \bibnamefont {Cheong}},\
  }\href {\doibase 10.1103/PhysRevB.64.115113} {\bibfield  {journal} {\bibinfo
  {journal} {Phys. Rev. B}\ }\textbf {\bibinfo {volume} {64}},\ \bibinfo
  {pages} {115113} (\bibinfo {year} {2001})}\BibitemShut {NoStop}%
\bibitem [{\citenamefont {Kolek}\ \emph {et~al.}(2007)\citenamefont {Kolek},
  \citenamefont {Stadler}, \citenamefont {Ptak}, \citenamefont {Zawislak},
  \citenamefont {Mleczko}, \citenamefont {Szalanski},\ and\ \citenamefont
  {Zak}}]{Kolek2007}%
  \BibitemOpen
  \bibfield  {author} {\bibinfo {author} {\bibfnamefont {A.}~\bibnamefont
  {Kolek}}, \bibinfo {author} {\bibfnamefont {A.~W.}\ \bibnamefont {Stadler}},
  \bibinfo {author} {\bibfnamefont {P.}~\bibnamefont {Ptak}}, \bibinfo {author}
  {\bibfnamefont {Z.}~\bibnamefont {Zawislak}}, \bibinfo {author}
  {\bibfnamefont {K.}~\bibnamefont {Mleczko}}, \bibinfo {author} {\bibfnamefont
  {P.}~\bibnamefont {Szalanski}}, \ and\ \bibinfo {author} {\bibfnamefont
  {D.}~\bibnamefont {Zak}},\ }\href {\doibase 10.1063/1.2815677} {\bibfield
  {journal} {\bibinfo  {journal} {Journal of Applied Physics}\ }\textbf
  {\bibinfo {volume} {102}},\ \bibinfo {eid} {103718} (\bibinfo {year}
  {2007})}\BibitemShut {NoStop}%
\bibitem [{\citenamefont {Mantese}\ \emph {et~al.}(1981)\citenamefont
  {Mantese}, \citenamefont {Goldburg}, \citenamefont {Darling}, \citenamefont
  {Craighead}, \citenamefont {Gibson}, \citenamefont {Buhrman},\ and\
  \citenamefont {Webb}}]{Mantese1981}%
  \BibitemOpen
  \bibfield  {author} {\bibinfo {author} {\bibfnamefont {J.}~\bibnamefont
  {Mantese}}, \bibinfo {author} {\bibfnamefont {W.}~\bibnamefont {Goldburg}},
  \bibinfo {author} {\bibfnamefont {D.}~\bibnamefont {Darling}}, \bibinfo
  {author} {\bibfnamefont {H.}~\bibnamefont {Craighead}}, \bibinfo {author}
  {\bibfnamefont {U.}~\bibnamefont {Gibson}}, \bibinfo {author} {\bibfnamefont
  {R.}~\bibnamefont {Buhrman}}, \ and\ \bibinfo {author} {\bibfnamefont
  {W.}~\bibnamefont {Webb}},\ }\href {\doibase 10.1016/0038-1098(81)90375-6}
  {\bibfield  {journal} {\bibinfo  {journal} {Solid State Communications}\
  }\textbf {\bibinfo {volume} {37}},\ \bibinfo {pages} {353 } (\bibinfo {year}
  {1981})}\BibitemShut {NoStop}%
\bibitem [{\citenamefont {Sampietro}\ \emph {et~al.}(1999)\citenamefont
  {Sampietro}, \citenamefont {Fasoli},\ and\ \citenamefont
  {Ferrari}}]{Sampietro1999}%
  \BibitemOpen
  \bibfield  {author} {\bibinfo {author} {\bibfnamefont {M.}~\bibnamefont
  {Sampietro}}, \bibinfo {author} {\bibfnamefont {L.}~\bibnamefont {Fasoli}}, \
  and\ \bibinfo {author} {\bibfnamefont {G.}~\bibnamefont {Ferrari}},\ }\href
  {\doibase 10.1063/1.1149785} {\bibfield  {journal} {\bibinfo  {journal}
  {Review of Scientific Instruments}\ }\textbf {\bibinfo {volume} {70}},\
  \bibinfo {pages} {2520} (\bibinfo {year} {1999})}\BibitemShut {NoStop}%
\bibitem [{\citenamefont {Jonker}\ \emph {et~al.}(1999)\citenamefont {Jonker},
  \citenamefont {Briaire},\ and\ \citenamefont {Vandamme}}]{Jonker1999}%
  \BibitemOpen
  \bibfield  {author} {\bibinfo {author} {\bibfnamefont {R.}~\bibnamefont
  {Jonker}}, \bibinfo {author} {\bibfnamefont {J.}~\bibnamefont {Briaire}}, \
  and\ \bibinfo {author} {\bibfnamefont {L.}~\bibnamefont {Vandamme}},\ }\href
  {\doibase 10.1109/19.772210} {\bibfield  {journal} {\bibinfo  {journal}
  {Instrumentation and Measurement, IEEE Transactions on}\ }\textbf {\bibinfo
  {volume} {48}},\ \bibinfo {pages} {730 } (\bibinfo {year}
  {1999})}\BibitemShut {NoStop}%
\bibitem [{\citenamefont {DiCarlo}\ \emph {et~al.}(2006)\citenamefont
  {DiCarlo}, \citenamefont {Zhang}, \citenamefont {McClure}, \citenamefont
  {Marcus}, \citenamefont {Pfeiffer},\ and\ \citenamefont
  {West}}]{Dicarlo2006}%
  \BibitemOpen
  \bibfield  {author} {\bibinfo {author} {\bibfnamefont {L.}~\bibnamefont
  {DiCarlo}}, \bibinfo {author} {\bibfnamefont {Y.}~\bibnamefont {Zhang}},
  \bibinfo {author} {\bibfnamefont {D.~T.}\ \bibnamefont {McClure}}, \bibinfo
  {author} {\bibfnamefont {C.~M.}\ \bibnamefont {Marcus}}, \bibinfo {author}
  {\bibfnamefont {L.~N.}\ \bibnamefont {Pfeiffer}}, \ and\ \bibinfo {author}
  {\bibfnamefont {K.~W.}\ \bibnamefont {West}},\ }\href {\doibase
  10.1063/1.2221541} {\bibfield  {journal} {\bibinfo  {journal} {Review of
  Scientific Instruments}\ }\textbf {\bibinfo {volume} {77}},\ \bibinfo {eid}
  {073906} (\bibinfo {year} {2006})}\BibitemShut {NoStop}%
\bibitem [{511()}]{5113}%
  \BibitemOpen
  \href@noop {} {}\bibinfo {note}
  {\url{http://www.signalrecovery.com/Our-Products/Preamplifiers/5113.aspx},
  accessed 1 July 2014}\BibitemShut {NoStop}%
\bibitem [{NI()}]{NI}%
  \BibitemOpen
  \href@noop {} {}\bibinfo {note}
  {\url{http://sine.ni.com/nips/cds/view/p/lang/en/nid/209219}, accessed 1 July
  2014}\BibitemShut {NoStop}%
\bibitem [{P-C(2012)}]{P-C-Pattern}%
  \BibitemOpen
  \href {http://www.ni.com/white-paper/3023/en} {\emph {\bibinfo {title}
  {Application Design Patterns: Producer/Consumer}}},\ \bibinfo {organization}
  {National Instruments} (\bibinfo {year} {2012})\BibitemShut {NoStop}%
\bibitem [{\citenamefont {Dutta}\ and\ \citenamefont {Horn}(1981)}]{Dutta1981}%
  \BibitemOpen
  \bibfield  {author} {\bibinfo {author} {\bibfnamefont {P.}~\bibnamefont
  {Dutta}}\ and\ \bibinfo {author} {\bibfnamefont {P.~M.}\ \bibnamefont
  {Horn}},\ }\href {\doibase 10.1103/RevModPhys.53.497} {\bibfield  {journal}
  {\bibinfo  {journal} {Rev. Mod. Phys.}\ }\textbf {\bibinfo {volume} {53}},\
  \bibinfo {pages} {497} (\bibinfo {year} {1981})}\BibitemShut {NoStop}%
\bibitem [{\citenamefont {McWhorter}(1955)}]{McWhorter1955}%
  \BibitemOpen
  \bibfield  {author} {\bibinfo {author} {\bibfnamefont {A.~L.}\ \bibnamefont
  {McWhorter}},\ }\href {http://hdl.handle.net/1721.1/12061} {\emph {\bibinfo
  {title} {1/f noise and related surface effects in germanium (Thesis (Sc.
  D.))}}}\ (\bibinfo  {publisher} {Massachusetts Institute of Technology},\
  \bibinfo {year} {1955})\BibitemShut {NoStop}%
\bibitem [{\citenamefont {Takagi}\ \emph {et~al.}(1986)\citenamefont {Takagi},
  \citenamefont {Mizunami}, \citenamefont {Suzuki},\ and\ \citenamefont
  {Masuda}}]{Takagi1986}%
  \BibitemOpen
  \bibfield  {author} {\bibinfo {author} {\bibfnamefont {K.}~\bibnamefont
  {Takagi}}, \bibinfo {author} {\bibfnamefont {T.}~\bibnamefont {Mizunami}},
  \bibinfo {author} {\bibfnamefont {J.}~\bibnamefont {Suzuki}}, \ and\ \bibinfo
  {author} {\bibfnamefont {S.}~\bibnamefont {Masuda}},\ }\href {\doibase
  10.1109/TCHMT.1986.1136638} {\bibfield  {journal} {\bibinfo  {journal}
  {Components, Hybrids, and Manufacturing Technology, IEEE Transactions on}\
  }\textbf {\bibinfo {volume} {9}},\ \bibinfo {pages} {141 } (\bibinfo {year}
  {1986})}\BibitemShut {NoStop}%
\bibitem [{\citenamefont {Fleetwood}\ \emph {et~al.}(1984)\citenamefont
  {Fleetwood}, \citenamefont {Postel},\ and\ \citenamefont
  {Giordano}}]{Fleetwood1984}%
  \BibitemOpen
  \bibfield  {author} {\bibinfo {author} {\bibfnamefont {D.~M.}\ \bibnamefont
  {Fleetwood}}, \bibinfo {author} {\bibfnamefont {T.}~\bibnamefont {Postel}}, \
  and\ \bibinfo {author} {\bibfnamefont {N.}~\bibnamefont {Giordano}},\ }\href
  {\doibase 10.1063/1.333845} {\bibfield  {journal} {\bibinfo  {journal}
  {Journal of Applied Physics}\ }\textbf {\bibinfo {volume} {56}},\ \bibinfo
  {pages} {3256} (\bibinfo {year} {1984})}\BibitemShut {NoStop}%
\bibitem [{\citenamefont {Tinti}\ \emph {et~al.}(2000)\citenamefont {Tinti},
  \citenamefont {Sischka},\ and\ \citenamefont {Morton}}]{Agilent}%
  \BibitemOpen
  \bibfield  {author} {\bibinfo {author} {\bibfnamefont {R.}~\bibnamefont
  {Tinti}}, \bibinfo {author} {\bibfnamefont {F.}~\bibnamefont {Sischka}}, \
  and\ \bibinfo {author} {\bibfnamefont {C.}~\bibnamefont {Morton}},\ }\href
  {http://cp.literature.agilent.com/litweb/pdf/5989-9087EN.pdf} {\emph
  {\bibinfo {title} {Proposed System Solution for 1/f Noise Parameter
  Extraction}}},\ \bibinfo {organization} {Agilent Technologies} (\bibinfo
  {year} {2000})\BibitemShut {NoStop}%
\bibitem [{\citenamefont {Nyquist}(1928)}]{Nyquist1928}%
  \BibitemOpen
  \bibfield  {author} {\bibinfo {author} {\bibfnamefont {H.}~\bibnamefont
  {Nyquist}},\ }\href {\doibase 10.1103/PhysRev.32.110} {\bibfield  {journal}
  {\bibinfo  {journal} {Phys. Rev.}\ }\textbf {\bibinfo {volume} {32}},\
  \bibinfo {pages} {110} (\bibinfo {year} {1928})}\BibitemShut {NoStop}%
\bibitem [{Vis(2010)}]{VishayDale}%
  \BibitemOpen
  \href {http://www.vishay.com/docs/31025/erc.pdf} {\emph {\bibinfo {title}
  {Vishay Dale ERC (Military RNC/RNR) Metal Film Resistors Datasheet}}},\
  \bibinfo {organization} {Vishay} (\bibinfo {year} {2010})\BibitemShut
  {NoStop}%
\end{thebibliography}%

\end{document}